\PassOptionsToPackage{dvipsnames}{xcolor}
\documentclass[sigconf,nonacm]{acmart}

\newcommand{\TODO}[1]{\textbf{\color{red}[TODO: #1]}}
\newcommand{\NUM}[2]{\ifcsname num#1\endcsname\csname num#1\endcsname\else\TODO{#2}\fi}
\newcommand\numToolShare{37.7}  %
\newcommand\numWaitFailMedSeq{7.1}  %
\newcommand\numWaitFailMedOurs{2.6}  %
\newcommand\numWaitFilmMedSeq{35.5}  %
\newcommand\numWaitFilmMedOurs{15.5}  %

\usepackage{microtype}   %

\usepackage{xspace}
\usepackage{multirow}
\newcommand{\sys}{CodeReel\xspace}
\newcommand{\bench}{EditLoop\xspace}
\newcommand{\rebuild}{Rebuild\xspace}
\newcommand{\sequential}{Sequential\xspace}

\usepackage{algorithm}
\usepackage[noend]{algpseudocode}
\algrenewcommand\algorithmicrequire{\textbf{Input:}}
\algrenewcommand\algorithmicensure{\textbf{Output:}}
\newcounter{property}
\newenvironment{property}[1][]{\par\smallskip\refstepcounter{property}\noindent\textbf{Property~\theproperty}%
  \if\relax\detokenize{#1}\relax\else\ (#1)\fi.\ \itshape\ignorespaces}{\par\smallskip}
\makeatletter\AtBeginDocument{\def\theHALG@line{\thealgorithm.\arabic{ALG@line}}}\makeatother
\AtEndPreamble{\crefname{property}{Property}{Properties}\crefname{algorithm}{Alg.}{Algs.}\Crefname{algorithm}{Algorithm}{Algorithms}\crefname{ALG@line}{line}{lines}\crefname{equation}{Eq.}{Eqs.}}
\usepackage{pgfplots}
\pgfplotsset{compat=1.18}
\usetikzlibrary{arrows.meta,positioning,decorations.pathreplacing}
\definecolor{cA}{HTML}{2A78D6}
\definecolor{cB}{HTML}{EB6834}
\definecolor{cC}{HTML}{1BAF7A}
\definecolor{cD}{HTML}{EDA100}
\definecolor{cE}{HTML}{6250D6}
\definecolor{cGray}{HTML}{C3C2B7}
\definecolor{cInk}{HTML}{52514E}

\usepackage[capitalize]{cleveref}
\crefname{section}{Sec.}{Secs.}
\Crefname{section}{Section}{Sections}
\Crefname{table}{Table}{Tables}
\crefname{table}{Tab.}{Tabs.}
\title{CodeReel: Accelerating Code-to-Video Agents with Exact Frame Reuse and Blocking-Aware Scheduling}
\author{Junyi Shen}
\affiliation{\institution{National University of Singapore}\city{Singapore}\country{Singapore}}
\email{j1shen@comp.nus.edu.sg}
\author{Yao Lu}
\affiliation{\institution{National University of Singapore}\city{Singapore}\country{Singapore}}
\email{luyao@comp.nus.edu.sg}
\begin{document}
\begin{abstract}
Coding agents make videos by writing and refining programs in a build--inspect--edit loop. A finished film costs a
median of \$\NUM{CostPerSession}{} in model calls and \NUM{MinutesPerSession}{} minutes, over a third of that time in builds that block the agent.
We find that much of this time is avoidable, for two reasons. First, \NUM{SeenKeysPct}{}\% of the frames of a session's later films were drawn before, many shifted in time or inside an
edited scene. Second, half of the builds stop at a check, where the agent needs only the verdict, but which builds stop is unknown until the checks finish.
We present \sys, a training-free execution layer behind agents' existing commands.
\emph{Exact frame reuse} keys each frame by its drawing commands, regardless of its position in time, and reuses each
frame whose key recurs.
\emph{Blocking-aware scheduling} runs the checks first and starts the render alongside them, at idle priority, unless
the session's recent builds make a pass unlikely. Scheduling spends CPU to answer sooner and reuse returns it; exact keys let later versions use frames rendered for a failed one as the agent thinks.
We introduce \bench, a benchmark of replayable agent sessions whose films are checked against renders from
scratch.
On it, \sys cuts the agent's wait by \NUM{WaitCutS}{}\% against Rebuild, a warm pipeline with parallel
rendering, and by \NUM{SeqOursPct}{}\% against Sequential, today's step-by-step pipeline. On time-shifting edits its reuse renders \NUM{ShiftTabOursPct}{}\%, close to the \NUM{ShiftTabTruthPct}{}\% whose pixels are new, while Manim-style
caching renders \NUM{ShiftPerCallPct}{}\%. With real agents, the median wait per build is \NUM{RWait}{}$\times$
Rebuild's in Claude Code, on topics never used in development, and \NUM{CodexWait}{}$\times$ in Codex CLI. The gain holds
for five models and two frameworks.
\end{abstract}

\maketitle
\section{Introduction}
\label{sec:intro}
\input{figures/fig_teaser}

Coding agents such as Claude Code~\cite{claudecode} and Codex CLI~\cite{codexcli} produce videos by writing
programs~\cite{code2video,theoremexplain,preacher,omnimanim,reelsrsi,compass}. To make an explainer film, the agent writes scene scripts, a narration timeline and assets, and then refines them in a loop (\cref{fig:teaser}a). In each round it builds the program, inspects the errors or the film, and revises the sources. Some revisions fix its own defects, such as text over subjects or captions out of sync with the speech~\cite{code2video,omnimanim,reelsrsi}; others answer its user's change requests. Each build is a tool call the agent waits for. A build runs four steps in sequence, as current pipelines run theirs~\cite{code2video,theoremexplain}: it synthesizes the narration's voice, lays out the timeline, checks the whole film and, if the checks pass, renders every frame. It returns a verdict, pass or fail, with the film on a pass and the errors on a fail. In \NUM{RealSessions}{} recorded
sessions of Claude Code, a finished film takes a median of \NUM{BuildsPerSession}{} builds and costs \$\NUM{CostPerSession}{} in model calls and \NUM{MinutesPerSession}{} minutes. The builds take over a third of that time,
each blocking the agent for a median of \NUM{BuildMedS}{}\,s.
We find that much of this wait is avoidable (\cref{sec:analysis}). The first cause is redundancy: \NUM{SeenKeysPct}{}\% of the frames of a session's later films were drawn by an earlier version, yet they are rendered again. The second is a
scheduling mismatch: \NUM{FailShare}{}\% of builds stop at a check before any film is made. The agent waits for the verdict or the film depending on an outcome known only when the checks finish, yet the pipeline runs every build in one fixed order.

\par\noindent\textbf{Challenges.}
\emph{Redundancy.} An agent revises a film incrementally, so the frames an edit does not touch recur: \NUM{InPlacePct}{}\% of later films' frames at the same time in the previous film and \NUM{ShiftedPct}{}\% at another time (\cref{fig:teaser}b and \cref{fig:time}b). Three kinds of edit make recurring frames hard to identify before rendering. First, an edit to a file that scenes
share changes only some frames of the scenes that read it. But a script draws a whole scene from one clock, with no call, node or layer boundary inside it. Caches keyed by an animation call as in Manim~\cite{manim}, a scene~\cite{fourier}, a node or layer~\cite{nuke,aegpc,blinkpaint} or a
file~\cite{buildsalacarte,riker} therefore redraw those scenes whole. Second, a narration edit moves every later scene. An incremental renderer such as kamishibai keys a frame by its index~\cite{kamishibai} and so misses frames that did not change. Third, many edits change little of a frame, such as a few words of a caption. Approximate reuse, as in training-free diffusion caches~\cite{deepcache,teacache,adacache,fastercache} and in edit-time methods that recompute only the edited regions~\cite{sige,lazydiffusion,fisedit}, reuses results whose inputs or features change little. But \NUM{ChangedNearPct}{}\% of the changed frames keep an SSIM~\cite{wang2004ssim} above 0.99 to the frame they replace, so it can keep a \emph{stale} frame that still shows the old words.
\emph{Scheduling mismatch.} A failing build
should give every core to its checks, and a passing one should start its render at once. The checks and the render
both process the whole film on the same cores, so a render alongside the checks slows the verdict, and one after them
delays the film. The machine is idle while the agent thinks, but the version to build next is not yet written. Pipelines fix the order that completes the film~\cite{code2video,theoremexplain}, schedulers for agents schedule around tool calls~\cite{llmcompiler,conveyor,specactions,specbox}, and test prioritization
reorders the tests of a suite~\cite{rothermel2001apfd,prado2020ntr}. None orders the work inside a build by what the agent is blocked on.

\par\noindent\textbf{Our approach.}
We present \sys, a training-free, model-agnostic execution layer behind the commands agents already call, built on
two observations. First, video programs draw each frame from its time alone~\cite{remotionflicker}, so the commands a frame issues
determine its pixels. Evaluating them costs \NUM{KeyCpuMs}{}\,ms of CPU, against \NUM{ShotCoreMs}{}\,ms per frame for a full render. \emph{Exact frame reuse} therefore keys a frame by its drawing commands, regardless of its position in time, and reuses every frame whose key was rendered before. Second, a build call is synchronous, so the agent is blocked on the verdict until the checks
return, on the film once they pass, and on nothing once the build has answered. \emph{Blocking-aware scheduling} runs the
checks first and in parallel, and decides from the session's history and the machine's cores whether to render alongside them. In the agent's think time it works ahead, rendering a failed version's frames, since a stored frame serves any later
version that draws it. Neither technique changes the verdict or the film of a build from scratch.

On recorded sessions replayed on 32 cores, \sys cuts the agent's wait by \NUM{WaitCutS}{}\% against \rebuild, a pipeline with warm processes and
workers sized to the machine (\cref{fig:teaser}c shows one build). Against \sequential, that pipeline run step by step as today's agents run theirs~\cite{code2video,theoremexplain}, the cut is \NUM{SeqOursPct}{}\%. \sys answers failing builds in \NUM{WaitFailS}{}$\times$ \rebuild's time. Most of the reduction comes from scheduling, which spends CPU to answer sooner; reuse renders \NUM{FramesCutS}{}\% fewer frames, returns that CPU and helps most where cores are scarce. Real agents wait less as well,
with a median wait per build of \NUM{RWait}{}$\times$ \rebuild's in Claude Code, on topics never used in development,
and \NUM{CodexWait}{}$\times$ in Codex CLI.
Of the reuse schemes without stale frames, \sys renders the fewest frames. On time-shifting edits it renders \NUM{ShiftTabOursPct}{}\%, close to the \NUM{ShiftTabTruthPct}{}\% whose pixels are new, while Manim-style caching~\cite{manim} renders
\NUM{ShiftPerCallPct}{}\%. On programs of another video framework, Remotion~\cite{remotion}, the agent waits
\NUM{MainOursB}{}$\times$ as long under \sys as under Remotion's own renderer (\cref{sec:experiments}).
We contribute:
\begin{itemize}
\item \textbf{\bench}: a benchmark of replayable agent sessions, every delivered film checked against a render from
  scratch (\cref{sec:analysis}).
\item \textbf{Exact frame reuse}: frames keyed by their drawing commands, shift-invariant, with cheaper tests first, returning the CPU that scheduling spends (\cref{sec:reuse}).
\item \textbf{Blocking-aware scheduling}: a build ordered by what the agent is blocked on, spending CPU to answer sooner, with work ahead in its think time that exact keys let later versions keep (\cref{sec:verdict}).
\item \textbf{\sys}: a general execution layer that existing coding agents use unmodified, across video frameworks
  (\cref{sec:method}).
\end{itemize}

\section{The \bench benchmark}
\label{sec:analysis}
\input{figures/fig_gallery}
Evaluating an execution layer for code-to-video agents needs a record of what agents did, in a form any system can
replay. Code-to-video benchmarks grade the film~\cite{code2video,compass}, a video-editing benchmark adds the
agent's runtime and cost~\cite{timelinebench}, and traces of coding agents record tool durations or turn lengths to benchmark LLM serving~\cite{tracelab,agentreplay,agentperfbench}. None keeps each program version with the tool calls that built it.
\bench keeps both, so every system can replay the same calls in order, and it checks every delivered film against a render from scratch. We release the sessions with the replayer and its correctness gate.

\subsection{Sessions and tracks}
A \emph{session} records one agent making one film: every version it committed, every tool call in order with its
timing and outcome, the agent's think time before each call, and the class of edit each build follows.
The \emph{real track} has \NUM{RealSessions}{} sessions of Claude Code~\cite{claudecode} with Claude Opus~5.5~\cite{claudemodels} making
explainer films with cv.js, our Canvas~2D~\cite{canvas2d} runtime (\cref{fig:gallery}): \NUM{RealVersions}{} versions, \NUM{RealCalls}{} tool calls and \NUM{RealBuilds}{} builds. Its films run \NUM{FilmPTenS}{} to \NUM{FilmPNinetyS}{}\,s between the 10th and 90th percentiles and are narrated in Chinese and English. Of its builds, \NUM{EditFilesPct}{}\% follow edits of new or removed files,
\NUM{EditLinesPct}{}\% of narration lines, \NUM{EditZonePct}{}\% of one scene's script and \NUM{EditSharedPct}{}\% of shared
scripts or data. It holds every session
recorded over two weeks that built at least two versions. A scripted user issued every change request. Of the sessions, \NUM{HeldoutSessions}{} are on
topics that GPT-6.1~\cite{openaimodels} wrote for \bench, each with three rounds of change
requests that build on one another.
The real track is split by topic: the \NUM{DevSessions}{} sessions on topics used to tune any parameter
form the development split, and the other \NUM{TestSessions}{}, on \NUM{TestTopics}{} topics that include all of GPT-6.1's, the test split.
Timing replays one session from each of \NUM{TimedSessions}{} topics, drawn with a fixed seed. The reuse study takes the \NUM{Pairs}{} pairs of adjacent delivered versions of \NUM{PairSessions}{} test-split sessions.
The \emph{model track} has \NUM{ModelSessions}{} sessions of Claude Sonnet~5.5 and Haiku~5.5~\cite{claudemodels} on \NUM{ModelTopics}{} test topics. The
\emph{Remotion track} brings a second framework, Remotion~\cite{remotion}, with synthetic edits whose classes and think times follow the real track's:
\NUM{RxVersions}{} versions in \NUM{RxSessions}{} sessions.

\subsection{Protocol and metrics}
Replay is closed-loop~\cite{schroeder2006openclosed}: a system executes a session's recorded calls in order, each after the previous one has answered, either at once or after the recorded think time. Caches start cold in every session.
\par\noindent\textbf{Correctness gate.} \bench requires the \emph{from-scratch consistency} of incremental computation~\cite{hammer2015adapton}, and reports a system's times only if every build keeps it. A build must return the verdict of a build from scratch and no error that the latter does not report, and every frame of every delivered film must equal a render of that version from scratch (supplementary material).
\par\noindent\textbf{Metrics.} Renderers are compared by render time~\cite{kamishibai}, which misses what blocks the agent. We therefore measure the \emph{wait}, from a build's call until the build answers the agent: at the verdict when it fails, after the film when it passes.

\subsection{Where the time goes}
\input{figures/fig_time}
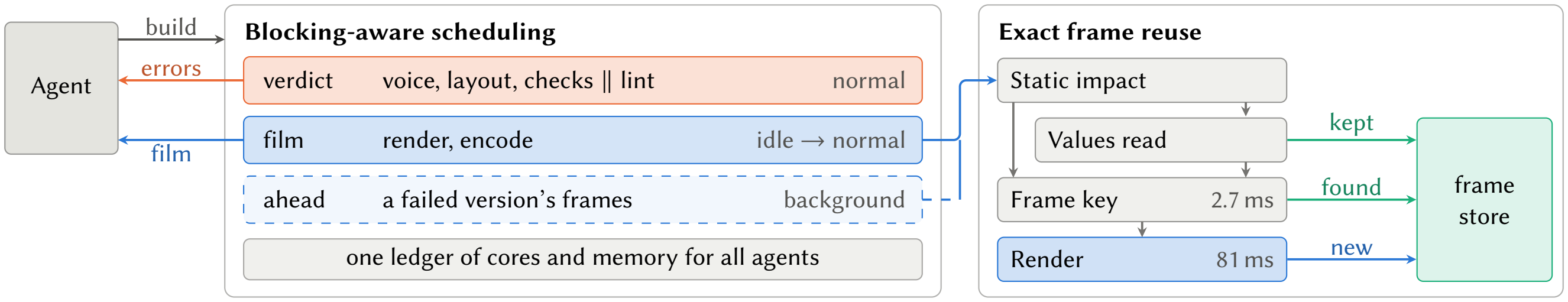
\begin{figure*}[t]
\centering
\begin{tikzpicture}[x=1pt, y=1pt, font=\sffamily\small,
  box/.style={rounded corners=2pt, inner xsep=4pt, inner ysep=2pt},
  arr/.style={-{Stealth[length=3.2pt, width=3pt]}, line width=0.7pt},
  part/.style={draw=cInk!45, rounded corners=3pt, fill=white},
  ttl/.style={inner sep=0, anchor=base west, font=\sffamily\small\bfseries},
  note/.style={inner sep=1pt, text=cInk},
  txt/.style={inner sep=0, anchor=west, text height=7pt, text depth=2pt},
  pri/.style={inner sep=0, anchor=east, text=cInk, text height=7pt, text depth=2pt},
  lane/.style={box, minimum height=15pt, minimum width=216pt},
  stage/.style={box, draw=cInk!60, fill=cGray!25, minimum height=14pt, minimum width=92pt}]
\node[box, draw=cInk!60, fill=cGray!45, minimum width=36pt, minimum height=42pt] (agent) at (18,71.5) {Agent};
\draw[part] (70,5) rectangle (298,98);
\node[ttl] at (76,87) {Blocking-aware scheduling};
\node[lane, draw=cB, fill=cB!20] at (184,74) {};
\node[txt] at (82,74) {verdict};
\node[txt] at (120,74) {voice, layout, checks $\parallel$ lint};
\node[pri] at (287,74) {normal};
\node[lane, draw=cA, fill=cA!20] at (184,55) {};
\node[txt] at (82,55) {film};
\node[txt] at (120,55) {render, encode};
\node[pri] at (287,55) {idle $\to$ normal};
\node[lane, draw=cA, dashed, fill=cA!6] at (184,36) {};
\node[txt] at (82,36) {ahead};
\node[txt] at (120,36) {a failed version's frames};
\node[pri] at (287,36) {background};
\node[box, draw=cInk!50, fill=cGray!25, minimum width=216pt, minimum height=13pt] at (184,17)
  {one ledger of cores and memory for all agents};
\draw[arr, cInk] (36,87) -- node[note, above] {build} (70,87);
\draw[arr, cB] (76,74) -- node[note, above, pos=0.575, text=cB!80!black] {errors} (36,74);
\draw[arr, cA] (76,55) -- node[note, below, pos=0.575, text=cA!80!black] {film} (36,55);
\draw[part] (310,5) rectangle (496,98);
\node[ttl] at (316,87) {Exact frame reuse};
\node[stage] (s1) at (362,74) {};
\node[txt] at ([xshift=4pt]s1.west) {Static impact};
\node[stage, minimum width=80pt] (s2) at (368,55) {};
\node[txt] at ([xshift=4pt]s2.west) {Values read};
\node[stage] (s3) at (362,36) {};
\node[txt] at ([xshift=4pt]s3.west) {Frame key};
\node[pri] at ([xshift=-4pt]s3.east) {\NUM{KeyCpuMs}{}\,ms};
\node[stage, draw=cA, fill=cA!22] (s4) at (362,17) {};
\node[txt] at ([xshift=4pt]s4.west) {Render};
\node[pri] at ([xshift=-4pt]s4.east) {\NUM{ShotCoreMs}{}\,ms};
\draw[arr, cInk!80] (321,67) -- (321,43);
\draw[arr, cInk!80] (395,67) -- (395,62.5);
\draw[arr, cInk!80] (395,48) -- (395,43);
\draw[arr, cInk!80] (362,29) -- (362,24);
\node[box, draw=cC, fill=cC!15, minimum width=43pt, minimum height=52pt, align=center] (store) at (471,36)
  {frame\\store};
\draw[arr, cC] (s2.east) -- node[note, above, text=cC!75!black] {kept} (s2.east -| store.west);
\draw[arr, cC] (s3.east) -- node[note, above, text=cC!75!black] {found} (s3.east -| store.west);
\draw[arr, cA] (s4.east) -- node[note, above, text=cA!80!black] {new} (s4.east -| store.west);
\draw[cA, line width=0.7pt, dashed] (292,36) -- (304,36);
\draw[arr, cA, rounded corners=2pt] (292,55) -- (304,55) -- (304,74) -- (s1.west);
\draw[cA, line width=0.7pt] (304,36) -- (304,55);
\end{tikzpicture}%
\vspace{-1mm}
\caption{\textbf{\sys.} The agent's synchronous build call enters the scheduler (\cref{alg:build}), which runs the
verdict's operations first, the film's alongside or after them, and a failed version's frames while the agent thinks.
Every render passes the reuse cascade, cheapest test first: static impact sends the frames an edit can reach to the
frame key, values read keeps the keys of the others, and only frames with new keys or no key are rendered, the keyed
ones into the frame store. Costs are CPU time per frame, including sound and encoding for the render, and per build for static impact.}
\label{fig:arch}
\end{figure*}
The real track yields three findings (\cref{fig:time}).

\noindent\textbf{F1: builds take over a third of a session.}
In the median session, tool calls take \NUM{ToolShare}{}\% of the time and builds alone \NUM{BuildShare}{}\%. Builds take a median of \NUM{BuildMedS}{}\,s and think times \NUM{TurnMedS}{}\,s.

\noindent\textbf{F2: half of the builds fail.}
\NUM{FailShare}{}\% of builds stop at a check, for text over subjects, clipped text, captions out of sync or lint
errors. In these builds the agent waits only for the verdict; in the others, the render waits for the checks.

\noindent\textbf{F3: later films repeat \NUM{SeenKeysPct}{}\% of their frames, unevenly.}
Over the \NUM{OVersions}{} films after a session's first in the timed test sessions,
\NUM{SeenKeysPct}{}\% of the frames repeat, command for command, a frame an earlier film drew. Of all frames, \NUM{InPlacePct}{}\% repeat at the
same time in the previous film, \NUM{ShiftedPct}{}\% at another time in it, and \NUM{HistoryPct}{}\% only in an older film. Of these films, \NUM{NewOverNinety}{}\% are over nine
tenths new and \NUM{NewUnderTen}{}\% under a tenth.

Exact reuse removes what F3 repeats; blocking-aware scheduling puts F2's checks first, overlaps the
render with them unless a pass is unlikely, and works in F1's think time.

\section{Method}
\label{sec:method}

\sys reduces the agent's wait in two complementary ways, one per observation of \cref{sec:intro} (\cref{fig:arch}): exact reuse across versions removes work and frees CPU, and blocking-aware scheduling spends it on what the agent waits for.

\subsection{Model}
\label{sec:model}
\par\noindent\textbf{Video programs.}\label{sec:contract}
A video program $P$ defines a film $F_P$, one frame $F_P(t)$ at each time $t$ of a frame grid. It draws each frame onto a canvas, or lays it out as a page, from $t$, its source files and the assets they load. \sys relies on the \emph{frame contract} Remotion states
for its programs~\cite{remotionflicker}: $F_P(t)$ depends on nothing else, not on randomness, earlier frames or wall-clock time. The runtime and the checks catch programs that break it (supplementary material).

\par\noindent\textbf{Builds.}
As build systems model a build~\cite{buildsalacarte}, a build of $P$ is a directed acyclic graph $G_P$ of CPU
operations, each serving a consumer. The verdict $\mathsf V$ needs each changed narration line voiced, the timeline laid out, the film checked at sampled times and the sources linted. The film $\mathsf F$ needs every frame and the sound rendered, the sound's levels checked and each segment encoded. A later version is the consumer of frames rendered ahead. The verdict $o_P\in\{\mathrm{pass},\mathrm{fail}\}$ is known at $T_{\mathsf V}$, when its checks
finish. A failing build answers with its errors at $T_{\mathsf V}$. A passing one answers at $T_{\mathsf F}\ge T_{\mathsf V}$, once its film and the checks that need the film have finished, with the film or those checks' errors. With times measured from the call and $[\cdot]$ the indicator, the agent waits
\begin{equation}
  W_P = T_{\mathsf V}\,[o_P=\mathrm{fail}] + T_{\mathsf F}\,[o_P=\mathrm{pass}].
  \label{eq:wait}
\end{equation}

\par\noindent\textbf{Problem.}
A session builds versions $P_1,\dots,P_n$, each given what the builds before it computed. The goal is to minimize the
expected total wait $\mathbb E\bigl[\sum_i W_{P_i}\bigr]$ over the outcomes, each unknown when its build starts.
\emph{Exactness} asks for the verdict
of a build of $P_i$ from scratch and, when it passes, for its film, frame by frame. \emph{Capacity} asks that the
operations running at once fit the machine's cores and memory, which other agents may share (supplementary material). \sys
removes from $G_{P_i}$ the operations already computed, shortening $T_{\mathsf V}$ and $T_{\mathsf F}$
(\cref{sec:reuse}), and orders the rest by consumer (\cref{sec:verdict}).

\subsection{Exact reuse across versions}
\label{sec:reuse}
Reuse must decide, without rendering a frame, whether its pixels can differ from those of a frame rendered before.
Following the first observation, \sys keys each frame by its drawing commands. Each part that draws the
frame at time $t$ of $P$, a scene's drawing or an overlay such as subtitles, records the commands it issues through the
runtime's supported interfaces, with their arguments and drawing state; images and fonts enter by their bytes. The frame's key is
\begin{equation}
  \kappa_P(t) = h\bigl(\sigma_P(t),\nu(a_1),\dots,\nu(a_m)\bigr),
  \label{eq:key}
\end{equation}
where $\sigma_P(t)$ holds the frame's structural inputs, such as its size, fonts and the order of its parts,
$a_1,\dots,a_m$ are the commands in drawing order, $\nu$ is their normalization to the rasterizer and $h$ is a hash.
Like the keys of build systems~\cite{buildsalacarte} and compositors~\cite{nuke}, a key is \emph{sound} if equal keys
imply equal outputs, so every reused output is exact, and \emph{tight} if equal outputs get equal keys, so no output is computed twice.
\begin{property}[Soundness]
\label{prop:sound}
For a frame drawn under the frame contract through supported interfaces, on one renderer configuration whose rasterizer draws a frame from its commands alone, with $\nu$ merging no two commands that the rasterizer draws differently and $h$ collision-free: if its key equals
that of a frame rendered before, its pixels equal that frame's.
\end{property}
A frame drawn through an unsupported interface gets no key and is rendered every time. \NUM{AuditKeyedPct}{}\% of the frames are keyed, and \cref{sec:exp-exact} reports the audit of every key.

\par\noindent\textbf{Normalizing keys to the rasterizer.}\label{sec:normalize}
If a key changes while the pixels do not, the frame is rendered needlessly; if it stays while they change, a stale
frame is shown. The normalization $\nu$ therefore rewrites only the commands whose rasterization we characterized. For example, the rasterizer places text at a baseline snapped to whole device pixels, so such text enters the key with the snapped baseline. Every other number enters a key exactly.

\par\noindent\textbf{Following frames through time.}\label{sec:shift}
A narration edit changes one line's duration and moves every later scene, so a cache looked up by time or frame
index~\cite{kamishibai} misses every frame of a moved scene, although none changed. The video framework Fourier follows whole layers that
move~\cite{fourier}. $\kappa$ is \emph{shift-invariant}, since it holds no absolute time: scenes draw from their local time and
start on whole frames. A scene that the next version $P'$ moves $\Delta$ frames later, its other inputs unchanged, therefore issues at $t+\Delta$ the
commands it issued at $t$, and $\kappa_{P'}(t+\Delta)=\kappa_P(t)$ for a frame whose other parts do the same.

\par\noindent\textbf{The cascade.}
Evaluating $\kappa$ for every frame would still cost a pass over the program, while the frames an edit cannot reach need
no new key, so cheaper tests come first (\cref{fig:arch}, right). \emph{Static impact} works like a build system's file
dependencies~\cite{buildsalacarte,riker}: a scene's script draws only inside the scene's time window, so the frames
inside the windows $\Omega$ that the changed scripts reach go to evaluation. \emph{Values read} tracks dependencies as
they are used, as incremental computation does~\cite{hammer2014adapton}: the runtime records hashes of the values each
frame read, such as the timeline, the style or an asset entry. A frame outside $\Omega$ keeps its key when they are
unchanged, so editing one style entry sends to evaluation only the frames that read it. Such a frame issues the same commands as before, so its kept key equals $\kappa_P(t)$ (Property~S1 of the supplementary material). For the remaining frames, evaluating $\kappa$ runs
the drawing code and records its commands, as browsers and UI toolkits keep display lists~\cite{blinkpaint,flutter}, without capturing
the frame.

\par\noindent\textbf{What is reused.}
A \emph{frame store} holds the rendered frames by key, per renderer configuration: a build takes from it every frame whose key it holds and renders
the rest, each missing key once. The film's encoded segments, which hold pictures only, are reused by their frames' keys, and the checks' results wherever static impact and values read keep a key.

\par\noindent\textbf{Another framework.}\label{sec:remotion}
The key and the store carry over to any runtime that exposes what each frame draws and keeps the frame contract: in
Remotion~\cite{remotion} this is the laid-out page, its elements with their computed styles and resources by content,
and frames with new keys go to its own renderer.

\subsection{Blocking-aware scheduling}
\label{sec:verdict}
The usual pipeline runs every operation at one priority, in the order that completes the
film~\cite{code2video,theoremexplain}. Following the second observation, \sys schedules each
operation by the consumer that waits for its output: the operations the verdict needs run first, those only the film
needs run alongside them when that pays off, and work no one waits for yet runs in the agent's think time. Each operation also takes its
consumer's share of the processor. The verdict's operations (\cref{alg:build},
\cref{alg:layout,alg:verdict}), and the film's once the agent waits for it (\cref{alg:film2}), run at normal priority. A render started early (\cref{alg:start}) runs at idle priority, which gets only the
processor time the others leave. Work ahead $A$ runs in the background, at low priority, until the next build
starts (\cref{alg:stop}) or blocking work needs memory, and keeps the frames it stored.
In \cref{alg:build}, \textsc{Voice}, \textsc{Layout}, \textsc{Check} and \textsc{Lint} are the verdict's operations of
\cref{sec:model}, and $\Lambda$ is the layout the checks and the render read. The operator $\parallel$ runs two operations at once, $o$ and $\mathcal E$ are the verdict and the errors, and \textsc{FilmCheck} runs the checks that need the film $F_P$. $\textsc{Render}(P,\Lambda,\pi)$ starts rendering $P$ in priority class $\pi$ and returns its handle $H$.

\begin{algorithm}[t]
\caption{One build of version $P$ on $c$ cores (\cref{sec:verdict}).}
\label{alg:build}
\small
\begin{algorithmic}[1]
\Require $P$, $\hat p$, $\theta_c$; work ahead $A$
\Ensure the answer to the agent
\State stop $A$; $H\gets\emptyset$ \label{alg:stop}
\State $e\gets[\hat p>\theta_c]$ \Comment{early render} \label{alg:early}
\State $\Lambda\gets\textsc{Layout}(\textsc{Voice}(P))$ \label{alg:layout}
\State \textbf{if} $e$ \textbf{then} $H\gets\Call{Render}{P,\Lambda,\mathrm{idle}}$ \Comment{asynchronous} \label{alg:start}
\State $(o,\mathcal E)\gets\textsc{Check}(P,\Lambda)\parallel\textsc{Lint}(P)$ \label{alg:verdict}
\If{$o=\mathrm{fail}$}
  \State answer $(o,\mathcal E)$; cancel $H$ \label{alg:fail}
  \State $A\gets\Call{Render}{P,\Lambda,\mathrm{background}}$ \Comment{work ahead} \label{alg:ahead}
\Else
  \State \textbf{if} $e$ \textbf{then} promote $H$ \textbf{else} $H\gets\Call{Render}{P,\Lambda,\mathrm{normal}}$ \label{alg:film2}
  \State wait for $H$; answer $F_P$, or the errors of $\Call{FilmCheck}{F_P}$
\EndIf
\end{algorithmic}
\end{algorithm}

\par\noindent\textbf{Within a build: the verdict first.}
On half of the builds the agent waits only for the verdict, whose checks cover the whole film. \sys runs the checks
first and in parallel, with the lint alongside them. Its \emph{fast checks} leave out what a verdict does not need: the checks that read a frame's texts and images sample frames as a key is evaluated, by running the drawing code without capturing pixels, and they map drawn objects to source lines only after the verdict. With few cores, a build also \emph{fails fast}, as in
test prioritization~\cite{rothermel2001apfd,prado2020ntr}: the checks that seldom fail run only once the others found
no error, and the build answers at its first errors.

\par\noindent\textbf{Within a build: the render as a bet.}
Starting the render alongside the checks is a bet on the outcome, like the sandboxes that agent serving
prewarms~\cite{specbox}. It saves render time when the version passes and slows the checks in either case, so it pays when the expected saving
exceeds the delay. At idle priority, the render slows the checks by a fraction $d$ and keeps a share $s$ of its speed, and the bet pays when the version passes with probability above $\theta_c=d/\bigl(s(1+d)\bigr)$
(supplementary material). Idle priority keeps the bet cheap, with a measured $\theta_c$ of \NUM{IdleNeedM}{} on 8 cores and \NUM{IdleNeedS}{} on 32. \Cref{alg:early} takes the bet when
$\hat p>\theta_c$, where $\hat p$ is the pass rate of the session's last five builds of the same kind, with the development split's as prior. A session's first build, which rarely passes,
thus renders after its verdict, and later builds render early (\cref{fig:teaser}c), \NUM{EarlyShareS}{}\% of builds on
32 cores and \NUM{EarlyShareM}{}\% on 8.

\par\noindent\textbf{Across the agent's think time.}
The machine would be idle while the agent reads a failed build's errors; serving systems use such windows too~\cite{mori}. The build therefore answers at its verdict, cancels any render started early
(\cref{alg:fail}) but keeps the frames it stored. It then renders the rest of the failed version's frames into the frame store
as work ahead (\cref{alg:ahead}), until the next build stops it. By \cref{prop:sound}, a stored frame is
exact for any later version that draws it.

\section{Experiments}
\label{sec:experiments}

\subsection{Experimental setup}
\par\noindent\textbf{Frameworks and hardware.}
Every system renders the same programs at 1080p and 30 frames per second: cv.js programs, written for our Canvas~2D runtime and run in headless Chromium, and Remotion~\cite{remotion} programs, written in React. The machines are 32, 8 or 4 of one server's cores, with 2\,GB of memory per core. 32 cores are abundant, since a render starts fewer workers than there
are cores; 8 are constrained, since a render can use more workers than cores; and 4 are saturated, since frame capture
alone fills them. Every system renders on the CPU (supplementary material).

\par\noindent\textbf{Baselines.}
For the wait, \rebuild is \sys without exact reuse and blocking-aware scheduling: one priority for every
operation, the checks in a single browser page with the lint after them, no fast checks or failing fast, and the render
after the checks. It keeps every other mechanism, such as warm browser pages, a resident voice process and workers sized to the machine. \sequential runs a
build's steps one after another on the same runtime, with cold browser pages, a cold voice process and eight render
workers, as code-to-video agents run theirs~\cite{code2video,theoremexplain}.
For reuse, we implement on the same films the schemes that tools and caches use. Scene windows re-render the time windows of the scenes whose files changed, and read sets re-render the frames whose files or read values changed. Kamishibai-style fingerprints~\cite{kamishibai} are \emph{typical}, over a part's file and timing, or \emph{careful}, adding every shared script and the whole timeline. Manim-style per-call caching keys each scene whole, by its files and the values it reads, as Manim keys each animation call~\cite{manim}. Complete keys reuse whole scenes or narration lines when all their frames' keys recur. The similarity rule reuses a frame as approximate caches do~\cite{teacache,nirvana}, on frame thumbnails.

\par\noindent\textbf{Protocol.}
\bench's test split is replayed as in \cref{sec:analysis}, each session once per system, back to back on the same cores in alternating order. Replays run without think time, which only work ahead uses (\cref{sec:exp-think}). The primary
estimate is the ratio of total waits over builds $b$, $\sum_b W_b/\sum_b W^{\mathrm{R}}_b$ with $W^{\mathrm{R}}_b$ the wait under \rebuild; 95\% intervals come from 2{,}000 bootstrap resamples of topics~\cite{efron1979bootstrap}. We registered the protocol before any formal run, with three claims: no
stale frame, a ratio below one on 8 and on 32 cores (Holm-corrected~\cite{holm1979}, one-sided 2.5\%), and a shorter median
wait per build for real agents.

\subsection{Comparison with existing systems}
\label{sec:exp-main}
\begin{figure}[t]
\centering
\newcommand\med[3]{\addplot[only marks, mark=*, mark size=1.5pt, color=cInk!70, forget plot] coordinates {(#1,0.5)};
  \addplot[only marks, mark=*, mark size=1.5pt, color=cA, forget plot] coordinates {(#2,0.5)};
  \node[anchor=north west, inner sep=1pt, font=\sffamily\small, text=cInk] at (axis cs:#1,0.5) {#1};
  \node[anchor=#3, inner sep=1pt, font=\sffamily\small, text=cA] at (axis cs:#2,0.5) {#2};}
\pgfplotsset{short legend/.style={legend image code/.code={\draw[##1] (0cm,0cm) -- (0.36cm,0cm);}}}
\begin{tikzpicture}
\pgfplotsset{wait/.style={
  scale only axis, width=82pt, height=54pt, ymin=0, ymax=1, xmin=0, clip=true,
  ytick={0,0.5,1}, table/col sep=comma, table/y=p,
  xlabel style={font=\sffamily\small, color=cInk}, ylabel style={font=\sffamily\small, color=cInk},
  ticklabel style={font=\sffamily\small, color=cInk}, axis line style={color=cGray}, tick style={color=cGray},
  axis x line*=bottom, axis y line*=left,
  seq/.style={const plot, no marks, color=cInk!70, densely dashed, line width=0.9pt},
  reb/.style={const plot, no marks, color=cGray, line width=1.3pt},
  ours/.style={const plot, no marks, color=cA, line width=1.3pt}}}
\begin{axis}[wait, name=fail, xmax=16, xtick={0,5,10,15}, xticklabels={0,5,10,15}, yticklabels={0,50\%,100\%},
  xlabel={time to verdict (s)}, ylabel={share of builds}, short legend,
  legend style={at={(1.09,1.06)}, anchor=south, draw=none, fill=none, font=\sffamily\small, legend columns=3, column sep=3pt},
  legend cell align=left]
\draw[cGray, dotted, line width=0.4pt] (axis cs:0,0.5) -- (axis cs:16,0.5);
\addplot[seq] table[x=serial] {figures/data/wait_fail.csv};
\addplot[reb] table[x=base] {figures/data/wait_fail.csv};
\addplot[ours] table[x=ours] {figures/data/wait_fail.csv};
\legend{\sequential, \rebuild, \sys}
\med{\numWaitFailMedSeq}{\numWaitFailMedOurs}{north west}
\end{axis}
\begin{axis}[wait, at={(fail.south east)}, anchor=south west, xshift=14pt, xmax=70, xtick={0,20,40,60},
  xticklabels={0,20,40,60}, yticklabels={}, xlabel={time to film (s)}]
\draw[cGray, dotted, line width=0.4pt] (axis cs:0,0.5) -- (axis cs:70,0.5);
\addplot[seq] table[x=serial] {figures/data/wait_film.csv};
\addplot[reb] table[x=base] {figures/data/wait_film.csv};
\addplot[ours] table[x=ours] {figures/data/wait_film.csv};
\med{\numWaitFilmMedSeq}{\numWaitFilmMedOurs}{south east}
\end{axis}
\end{tikzpicture}%
\vspace{-2mm}
\caption{\textbf{The agent's wait per build on 32 cores}, as the share of builds answered within a time, over the
\NUM{WaitFailBuilds}{} builds that end at the verdict and the \NUM{WaitFilmBuilds}{} that deliver a film; dots, the
medians (s).}
\label{fig:wait}
\end{figure}
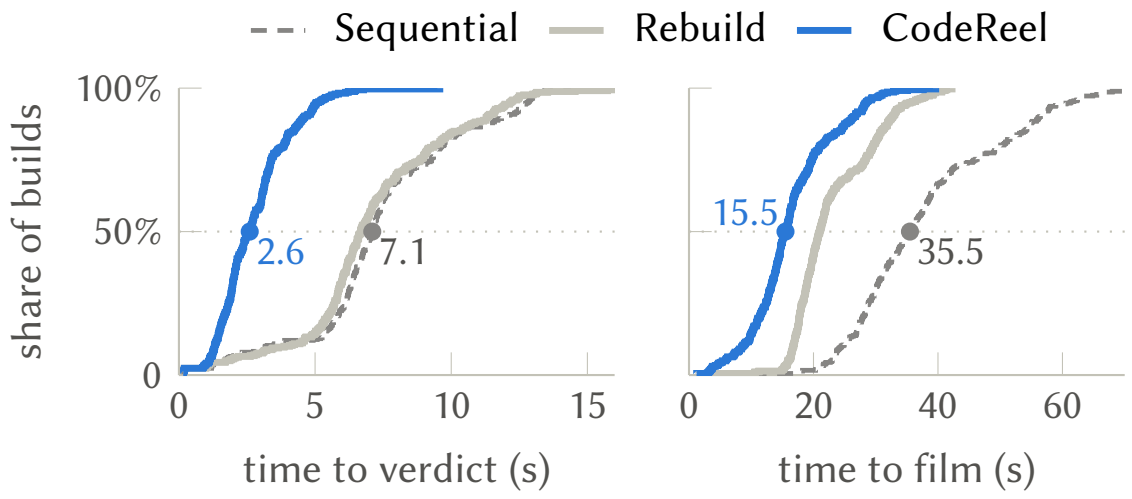

\begin{table}[t]
\centering\footnotesize\setlength{\tabcolsep}{2.6pt}
\caption{\textbf{\sys against existing systems by workload.} A block's first row gives the mean wait, CPU-seconds and
frames per build and the peak memory; the other rows are relative to it. Exact counts the delivered films equal to a render
from scratch (--: an external renderer, which renders each version from scratch).}
\label{tab:main}
\begin{tabular}{lrrrrr}
\toprule
System & Wait & CPU & Memory & Frames & Exact \\
\midrule
\multicolumn{6}{l}{\emph{EditLoop's real track (Canvas~2D, 32 cores; 70 topics)}} \\
\sequential~\cite{code2video,theoremexplain} & 23.2\,s & 143\,s & 4.3\,GB & 1{,}399 & 363/363 \\
\rebuild & 0.66 & 1.17 & 1.83 & 1.00 & 363/363 \\
\textbf{\sys} & \textbf{0.42} & \textbf{0.95} & \textbf{1.70} & \textbf{0.65} & \textbf{363/363} \\
\midrule
\multicolumn{6}{l}{\emph{EditLoop's Remotion track (React, 8 cores; synthetic edits)}} \\
Remotion renderer~\cite{remotion} & 11.9\,s & 91\,s & 2.2\,GB & 1{,}003 & -- \\
HyperFrames~\cite{hyperframes} & 0.96 & 0.59 & 0.93 & 1.00 & -- \\
\textbf{\sys} & \textbf{0.51} & \textbf{0.20} & \textbf{0.88} & \textbf{0.06} & \textbf{150/150} \\
\midrule
\multicolumn{6}{l}{\emph{kamishibai's edit benchmark~\cite{kamishibai} (React, 8 cores)}} \\
Remotion renderer~\cite{remotion} & 25.4\,s & 104\,s & 2.3\,GB & 2{,}580 & -- \\
HyperFrames~\cite{hyperframes} & 1.52 & 1.50 & 0.87 & 1.00 & -- \\
kamishibai, full~\cite{kamishibai} & 1.42 & 1.14 & 0.58 & 1.00 & -- \\
kamishibai, incremental~\cite{kamishibai} & 0.80 & 0.48 & 1.43 & 0.25 & 15/15 \\
\textbf{\sys} & \textbf{0.66} & \textbf{0.51} & \textbf{0.76} & \textbf{0.25} & \textbf{15/15} \\
\bottomrule
\end{tabular}

\end{table}
\Cref{tab:main} shows that \sys answers soonest on every workload. Each external system runs as shipped, on the programs written for it, with as many render workers as \sys. On \bench's real track, relative to \sequential, the agent waits \NUM{SeqOurs}{}$\times$ as long under \sys, which uses less CPU, and \NUM{SeqBase}{}$\times$ as long under \rebuild, which uses more. \Cref{fig:wait} gives the wait in seconds. \sys answers
\NUM{WaitFasterPct}{}\% of the builds sooner than \rebuild does. On the Remotion track, replayed on 8 cores, the agent waits
\NUM{MainOursB}{}$\times$ as long under \sys as under Remotion's own renderer~\cite{remotion}, with \NUM{MainOursFramesB}{}\% of the frames rendered. Under HyperFrames~\cite{hyperframes}, an HTML video framework for agents that renders each version from scratch, the agent waits \NUM{MainHfB}{}$\times$ as long on the programs ported to it. On kamishibai's own benchmark~\cite{kamishibai}, five edits of its reel after a
first render, \sys takes \NUM{MainOursOverKamiC}{}$\times$ the time of kamishibai run incrementally, which renders the
frames whose fingerprint changed.

\subsection{Wait across machines and agents}
\label{sec:exp-latency}
\begin{table}[t]
\centering\footnotesize\setlength{\tabcolsep}{4pt}
\caption{\textbf{Wait, frames and CPU relative to \rebuild.} \sys by machine, \rebuild with one mechanism on 32 and 8 cores,
and \sys without one on 32 cores, in a second replay whose first row is the full system.}
\label{tab:results}\label{tab:latency}\label{tab:ablation}
\begin{tabular}{lrrrrr}
\toprule
Setting & Wait & Failed & Film & Frames & CPU \\
\midrule
\multicolumn{6}{l}{\emph{\sys by machine}} \\
32 cores & 0.64 & 0.39 & 0.71 & 0.65 & 0.81 \\
32 cores, think time & 0.59 & 0.41 & 0.65 & 0.49 & 0.92 \\
8 cores & 0.67 & 0.43 & 0.71 & 0.66 & 0.76 \\
8 cores, think time & 0.60 & 0.43 & 0.63 & 0.53 & 0.86 \\
4 cores & 0.75 & 0.46 & 0.79 & 0.63 & 0.72 \\
\midrule
\multicolumn{6}{l}{\emph{\rebuild with one mechanism}} \\
32 cores, reuse & 0.94 & 0.97 & 0.94 & 0.66 & 0.81 \\
32 cores, verdict first & 0.69 & 0.49 & 0.75 & 1.00 & 1.12 \\
32 cores, fast checks & 0.82 & 0.58 & 0.89 & 1.00 & 0.98 \\
8 cores, reuse & 0.86 & 0.96 & 0.84 & 0.66 & 0.82 \\
8 cores, verdict first & 0.86 & 0.60 & 0.90 & 1.00 & 1.05 \\
8 cores, fast checks & 0.88 & 0.57 & 0.94 & 1.00 & 0.96 \\
\midrule
\multicolumn{6}{l}{\emph{\sys without one mechanism, on 32 cores}} \\
every mechanism & 0.62 & 0.38 & 0.69 & 0.65 & 0.77 \\
without reuse & 0.65 & 0.38 & 0.73 & 1.00 & 0.98 \\
without verdict first & 0.76 & 0.56 & 0.82 & 0.65 & 0.77 \\
without fast checks & 0.65 & 0.47 & 0.69 & 0.64 & 0.87 \\
never rendering early & 0.66 & 0.37 & 0.73 & 0.65 & 0.74 \\
\bottomrule
\end{tabular}

\end{table}
Under \sys the agent waits less than under \rebuild on every machine, for every model and in both harnesses,
Claude Code and Codex CLI, on failing and on delivering builds (Failed and Film in \cref{tab:latency}).

The registered tests on 32 and 8 cores replay the \NUM{TopicsS}{} timed topics, and the run on 4 cores replays
\NUM{TopicsW}{} of them. On 32 cores,
the main setting, the agent waits \NUM{WaitS}{}$\times$ as long (95\% CI \NUM{WaitCIS}{}):
\NUM{WaitFailS}{}$\times$ on failing builds and \NUM{WaitFilmS}{}$\times$ on builds that deliver a film. \sys renders \NUM{FramesCutS}{}\% fewer frames and, over all builds, returns the verdict in \NUM{VerdictS}{}$\times$ the time. On 8 cores the agent
waits \NUM{WaitM}{}$\times$ as long (CI \NUM{WaitCIM}{}), and on 4 cores \NUM{WaitW}{}$\times$ (CI \NUM{WaitCIW}{}). Both registered tests pass, and every build passes the correctness gate. On the \NUM{UntouchedSessions}{} live sessions of \cref{sec:exp-agents} on topics never used in development, the agent waits \NUM{UntouchedWaitS}{}$\times$ as long on 32 cores and \NUM{UntouchedWaitM}{}$\times$ on 8.

With the other models and harnesses, Claude Sonnet and Haiku from the model track, DeepSeek V4~\cite{deepseekv4} in Claude Code and GPT-6.1 in Codex CLI, the agent waits \NUM{OtherWaitLo}{}--\NUM{OtherWaitHi}{}$\times$ as long on 32 cores, each interval
below one.

\begin{figure}[t]
\centering
\pgfplotsset{agent/.style 2 args={x filter/.code={\edef\lawa{\thisrow{agent}}\edef\lawb{#1}\ifx\lawa\lawb\else\def\pgfmathresult{inf}\fi},
  only marks, mark size=#2, forget plot}}
\pgfplotsset{mark legend/.style={legend image code/.code={\draw[mark repeat=2, mark phase=2, ##1] plot coordinates {(0cm,0cm) (0.2cm,0cm) (0.4cm,0cm)};}}}
\begin{tikzpicture}
\pgfplotsset{lawpanel/.style={
  scale only axis, height=62pt, ymin=0, ymax=1.35, clip=false, ytick={0,0.5,1},
  xlabel style={font=\sffamily\small, color=cInk}, ylabel style={font=\sffamily\small, color=cInk},
  ticklabel style={font=\sffamily\small, color=cInk}, axis line style={color=cGray}, tick style={color=cGray},
  axis x line*=bottom, axis y line*=left, table/col sep=comma}}
\begin{axis}[lawpanel, name=law, width=84pt, xmin=0, xmax=1,
  xtick={0,0.5,1}, xticklabels={0,0.5,1}, yticklabels={0,0.5,1}, xlabel={frames rendered}, ylabel={wait, relative to \rebuild},
  table/x=frames, table/y=wait,
  legend style={at={(-0.36,-0.5)}, anchor=north west, draw=none, fill=none, font=\sffamily\small, legend columns=3,
    column sep=5pt, row sep=-1pt}, legend cell align=left,
  session/.style={only marks, forget plot},
]
\addlegendimage{only marks, mark=*, mark size=1.7pt, mark options={fill=cA, draw=white, line width=0.4pt}}
\addlegendimage{only marks, mark=*, mark size=1.5pt, mark options={fill=white, draw=cA, line width=0.7pt}}
\addlegendimage{only marks, mark=square*, mark size=1.5pt, mark options={fill=cB, draw=white, line width=0.4pt}}
\addlegendimage{only marks, mark=triangle*, mark size=2.1pt, mark options={fill=cB, draw=white, line width=0.4pt}}
\addlegendimage{only marks, mark=diamond*, mark size=2.1pt, mark options={fill=cC, draw=white, line width=0.4pt}}
\addlegendimage{only marks, mark=triangle*, mark size=2.1pt, mark options={fill=cC, draw=white, line width=0.4pt, rotate=180}}
\addplot[domain=0:1, samples=2, color=cInk, dashed, thin, forget plot] {x};
\def\lawfita{0.3754}\def\lawfitb{0.3711}
\addplot[domain=0:1, samples=2, color=cInk, thin, forget plot] {\lawfita + \lawfitb*x};
\addplot[session, mark=*, mark size=1.1pt, mark options={fill=cGray!80, draw=cGray!80, line width=0pt}] table {figures/data/law_opus.csv};
\addplot[session, mark=*, mark size=1.1pt, mark options={fill=cGray!80, draw=cGray!80, line width=0pt}] table {figures/data/law_claude.csv};
\addplot[session, mark=*, mark size=1.1pt, mark options={fill=cGray!80, draw=cGray!80, line width=0pt}] table {figures/data/law_sonnet.csv};
\addplot[session, mark=*, mark size=1.1pt, mark options={fill=cGray!80, draw=cGray!80, line width=0pt}] table {figures/data/law_haiku.csv};
\addplot[session, mark=*, mark size=1.1pt, mark options={fill=cGray!80, draw=cGray!80, line width=0pt}] table {figures/data/law_dpsk.csv};
\addplot[session, mark=*, mark size=1.1pt, mark options={fill=cGray!80, draw=cGray!80, line width=0pt}] table {figures/data/law_codex.csv};
\addplot[agent={opus}{2.7pt}, mark=*, mark options={fill=cA, draw=white, line width=0.7pt}] table {figures/data/law_pooled.csv};
\addplot[agent={claude}{2.5pt}, mark=*, mark options={fill=white, draw=cA, line width=1.1pt}] table {figures/data/law_pooled.csv};
\addplot[agent={sonnet}{2.4pt}, mark=square*, mark options={fill=cB, draw=white, line width=0.7pt}] table {figures/data/law_pooled.csv};
\addplot[agent={haiku}{3.3pt}, mark=triangle*, mark options={fill=cB, draw=white, line width=0.7pt}] table {figures/data/law_pooled.csv};
\addplot[agent={dpsk}{3.3pt}, mark=diamond*, mark options={fill=cC, draw=white, line width=0.7pt}] table {figures/data/law_pooled.csv};
\addplot[agent={codex}{3.3pt}, mark=triangle*, mark options={fill=cC, draw=white, line width=0.7pt, rotate=180}] table {figures/data/law_pooled.csv};
\legend{Opus 5.5, Opus 5.5$^\dagger$, Sonnet 5.5, Haiku 5.5, DeepSeek V4$^\dagger$, Codex CLI$^\dagger$}
\node[anchor=south west, inner sep=0pt, font=\sffamily\small\bfseries, text=black] at ([yshift=3pt]rel axis cs:0,1) {a};
\end{axis}
\pgfplotsset{panel/.style={
  scale only axis, width=70pt, height=62pt, table/col sep=comma, clip=false,
  xlabel style={font=\sffamily\small, color=cInk}, ylabel style={font=\sffamily\small, color=cInk},
  ticklabel style={font=\sffamily\small, color=cInk}, axis line style={color=cGray}, tick style={color=cGray},
  axis x line*=bottom, axis y line*=left}}
\begin{axis}[panel, name=cost, width=64pt, at={(law.south east)}, anchor=south west, xshift=37pt, xmin=0, xmax=1, ymin=0, ymax=1,
  xtick={0,0.5,1}, xticklabels={0,0.5,1}, ytick={0,0.5,1}, yticklabels={0,0.5,1},
  xlabel={frames with new pixels}, ylabel={frames rendered}, table/x=changed, pair/.style={only marks}, mark legend,
  legend style={at={(1,1.03)}, anchor=south east, draw=none, fill=none, font=\sffamily\small, inner sep=0pt}, legend cell align=left]
\addlegendimage{only marks, mark=square*, mark size=1.5pt, mark options={fill=cB, draw=white, line width=0.4pt}}
\addplot[pair, forget plot, mark=square*, mark size=1.1pt, mark options={fill=cB!85, draw=white, line width=0.2pt}] table[y=percall] {figures/data/cost.csv};
\addplot[pair, forget plot, mark=*, mark size=1.3pt, mark options={fill=cA, draw=white, line width=0.2pt}] table[y=ours] {figures/data/cost.csv};
\addplot[domain=0:1, samples=2, color=cInk, densely dashed, line width=0.7pt, forget plot] {x};
\node[anchor=south east, inner sep=1pt, font=\sffamily\small, text=cA] at (axis cs:0.99,0.03) {\sys};
\legend{Manim-style}
\node[anchor=south west, inner sep=0pt, font=\sffamily\small\bfseries, text=black] at ([yshift=3pt]rel axis cs:0,1) {b};
\end{axis}
\end{tikzpicture}%
\vspace{-2mm}
\caption{\textbf{The gain across sessions and edits.} (a) Wait against frames rendered, both relative to \rebuild, per session (gray)
and per agent (colored; $\dagger$: each agent's first session on nine topics); solid, the per-session fit; dashed, $y=x$. (b)
Frames rendered against frames with new pixels, per adjacent pair, \sys and Manim-style caching~\cite{manim}; points
below the dashed line, pixel truth, reuse stale frames.}
\label{fig:law}
\end{figure}
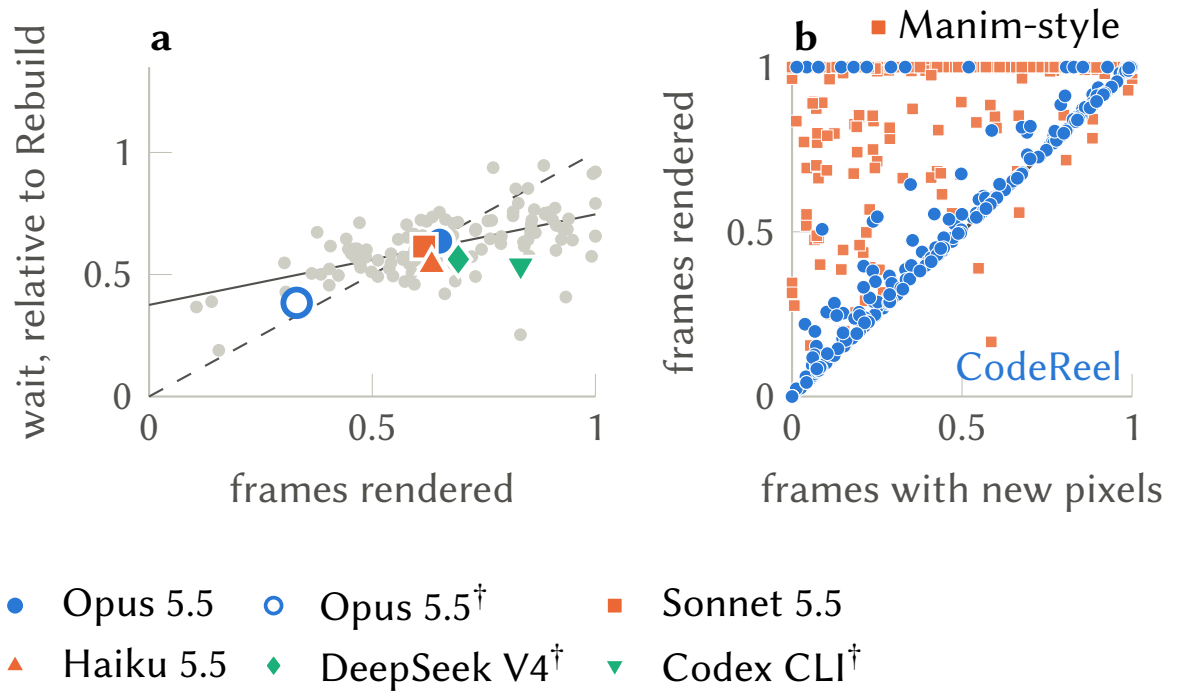
\par\noindent\textbf{Across sessions.}\label{sec:exp-law}
The agent waits less in sessions that render fewer frames again (\cref{fig:law}a): on the least-squares line,
$\text{wait}=\NUM{LawFitA}{}+\NUM{LawFitB}{}\cdot\text{frames}$, both relative to \rebuild. \Cref{sec:exp-ablation} separates what
reuse and scheduling each contribute.

\par\noindent\textbf{Think time and work ahead.}\label{sec:exp-think}
Replayed with the recorded think time, work ahead shortens the wait further, to \NUM{ThinkWaitL}{}$\times$ \rebuild's
on 8 cores and \NUM{ThinkWaitLS}{}$\times$ on 32 (\cref{tab:results}). On 8 cores that is \NUM{ThinkWaitPre}{}$\times$ the wait without work ahead.

\par\noindent\textbf{Shared machines.}
A machine is one budget for all its agents: \sys keeps a ledger of its cores and memory and sizes each render to what
is left, blocking work first, so every agent gets its films where uncoordinated renders outgrow the machine
(supplementary material).

\subsection{Exactness and tightness of reuse}
\label{sec:exp-exact}
\begin{table}[t]
\centering\footnotesize\setlength{\tabcolsep}{3.2pt}
\caption{\textbf{Frames rendered by each reuse scheme} on the test split's adjacent versions and on time-shifting
narration edits (cv.js). Exact: no stale frame; similarity: a frame reused when its
gray thumbnail differs from the previous version's by less than 0.1 on average; pixel truth: the frames
whose pixels are new.}
\label{tab:exact}
\begin{tabular}{lcrr}
\toprule
 & & Adjacent & Time-shifting \\
Method & Exact & versions & edits \\
\midrule
Scene windows & $\times$ & 90.3\% & 100.0\% \\
Read sets & $\times$ & 83.8\% & 53.2\% \\
Kamishibai-style, typical~\cite{kamishibai} & $\times$ & 63.5\% & 54.0\% \\
Kamishibai-style, careful & $\times$ & 89.0\% & 99.8\% \\
Manim-style~\cite{manim} & $\times$ & 88.0\% & 99.8\% \\
Similarity, threshold 0.1~\cite{teacache} & $\times$ & 59.7\% & -- \\
\addlinespace
Per scene, complete keys & $\checkmark$ & 80.2\% & 81.1\% \\
Per line, complete keys & $\checkmark$ & 71.1\% & 64.5\% \\
\textbf{\sys} & $\checkmark$ & \textbf{63.2\%} & \textbf{43.9\%} \\
\midrule
Pixel truth & $\checkmark$ & 60.2\% & 43.0\% \\
\bottomrule
\end{tabular}

\end{table}
Over the \NUM{PairFrames}{} frames of the reuse study's \NUM{Pairs}{} pairs, \sys reuses no frame whose pixels changed and
renders \NUM{RenderedPct}{}\% of the frames, \NUM{Tightness}{}$\times$ those whose pixels are new. Every other scheme renders more or reuses stale frames (\cref{tab:exact}).  The similarity rule of
\cref{tab:exact} leaves
\NUM{SimStrictStalePct}{}\% of all frames stale. These frames miss the agent's edit (\cref{fig:gallery}, bottom) at a median SSIM of \NUM{StaleSsim}{} to the correct frame (supplementary material). No frame \sys reuses is
stale, here or in \NUM{RandEdits}{} random edits, and an audit of every key against a lossless capture of its frame
finds no key that stands for two pictures (supplementary material).

\par\noindent\textbf{The cost of each edit.}\label{sec:exp-cost}
On \NUM{EditWithinFivePct}{}\% of pairs, \sys stays within 5 points of the share of frames whose pixels are new
(\cref{fig:law}b), while Manim-style caching renders whole
scenes.

\par\noindent\textbf{Time-shifting edits.}\label{sec:exp-shift}
In \NUM{ShiftEdits}{} narration edits to four cv.js films (\cref{tab:exact},
right), \sys renders \NUM{ShiftTabOursPct}{}\% of their frames, against \NUM{ShiftTabTruthPct}{}\% whose
pixels are new, and reuses no stale frame. Read sets reuse \NUM{ShiftTabReadStale}{} stale frames. Remotion films give similar results (supplementary material).

\subsection{Real agents in the loop}
\label{sec:exp-agents}
\begin{table}[t]
\centering\footnotesize\setlength{\tabcolsep}{1.9pt}
\caption{\textbf{Real agents on 8 cores}, \sys relative to \rebuild (geometric means over pairs; Faults: \sys minus
\rebuild, every 95\% interval including zero).}
\label{tab:agents}
\begin{tabular}{lrrrrrrr}
\toprule
Agent & Pairs & Wait & Film & Total & Builds & Time & Faults \\
\midrule
Claude Code, untouched & 14 & 0.62 & 0.76 & 0.69 & 0.93 & 0.83 & $-$0.29 \\
Claude Code, all & 26 & 0.59 & 0.73 & 0.70 & 1.01 & 0.84 & $-$0.42 \\
Codex CLI & 12 & 0.67 & 0.58 & 0.68 & 0.98 & 0.83 & $-$0.58 \\
\bottomrule
\end{tabular}

\end{table}
Under \sys, both agents wait less per build (\cref{tab:agents}). Each agent runs with the same model, topics and
scripted user under \rebuild and under \sys. Claude Code runs with Claude Opus~5.5 in 26 pairs of sessions on 16 topics, six of them from Code2Video's benchmark~\cite{code2video}; Codex CLI runs with GPT-6.1 on those six topics, twice each. Each pair's ratio compares the two sessions' median wait per build (Wait) and per delivering build (Film), total
wait (Total), builds and wall-clock time (Time). Faults is the difference in the number of defects that Claude
Opus~5.5~\cite{claudemodels}, blind to the system, lists in the final film. On the \NUM{RPairs}{} untouched pairs, on topics never used in development, the registered median wait per build drops to \NUM{RWait}{}$\times$ (CI \NUM{RWaitCI}{}) and the total wait to
\NUM{AgentTotalUntouched}{}$\times$. Through Codex CLI, the wait per build drops to \NUM{CodexWait}{}$\times$ (CI
\NUM{CodexWaitCI}{}) and the total to \NUM{AgentTotalCodex}{}$\times$. A session ends sooner as well (Time). The films' faults do not differ, and every final film \sys delivered equals a render from scratch.

\subsection{Ablation study}
\label{sec:exp-ablation}
As \cref{tab:results} shows, most of the reduction on 32 cores comes from scheduling, which spends CPU to answer sooner, while reuse cuts frames, returns that CPU and helps most where cores are scarce. Added alone to \rebuild, reuse renders \NUM{AblReuseFrames}{}\% fewer frames and
shortens the wait to \NUM{AblReuse}{}$\times$ on 32 cores but \NUM{AblReuseM}{}$\times$ on 8, where rendering fills the machine. The verdict first alone, with the checks in parallel and the render alongside them, shortens it to
\NUM{AblSched}{}$\times$ on 32 cores with \NUM{AblSchedCpu}{}$\times$ the CPU. Removing any one mechanism from \sys lengthens the wait, most of all the verdict first (to \NUM{AblNoVerdict}{}$\times$). Without reuse the CPU rises to
\NUM{AblNoReuseCpu}{}$\times$.

\section{Related work}
\label{sec:related}
\noindent\textbf{Code-to-video agents.} Agents write Manim, web or slide programs refined in a render
loop~\cite{code2video,theoremexplain,omnimanim,reelsrsi,paper2video}, check animations in that
loop~\cite{mover,logomotion}, drive simulators~\cite{moregen} or mix programs with generative models~\cite{preacher}. None reuses
more than its framework's per-call cache, and none targets the agent's wait.

\noindent\textbf{Reuse.} Training-free caches~\cite{deepcache,teacache,taylorseer}, edit-time methods~\cite{sige,lazydiffusion,hetcache}, render caches~\cite{rendercache,reprojection} and video networks
that skip unchanged regions~\cite{skipconv,deltacnn,eventful} approximate the full computation. Production tools,
browsers and UI frameworks cache or skip by node, layer, object or component~\cite{nuke,aegpc,blinkpaint,flutter,reactmemo,composeskip}. Closest to us, Manim keys animation
calls by their inputs~\cite{manim}, kamishibai fingerprints a frame's page at the same index~\cite{kamishibai},
HyperFrames repeats frames its timeline predicts static~\cite{hyperframes}, and Fourier fingerprints scenes and layers,
also after a layer moves in time~\cite{fourier}. Browsers and UI toolkits keep drawing commands as display lists within one page's lifetime~\cite{blinkpaint,flutter}; \sys keys whole frames by their evaluated commands, so frames of imperative code with one clock are found in any version at any time (\cref{tab:exact}). Like build systems and their content-addressed caches~\cite{buildsalacarte,riker,bazelcache,hammer2014adapton}, it is consistent with a build from scratch, but it keys what a program draws instead of what it reads.

\noindent\textbf{Scheduling around agents.} Test prioritization runs likely failures
first~\cite{rothermel2001apfd,prado2020ntr}, and agent systems schedule around tool
calls~\cite{llmcompiler,conveyor,specactions,mori,askthetool,specbox,agentcgroup}. \sys schedules within a
build by what the agent is blocked on, and the frames it renders ahead need no rollback.

\section{Limitations and conclusion}
\label{sec:conclusion}
\sys shortens a code-to-video agent's wait by exact frame reuse, which removes repeated work, and blocking-aware scheduling, which spends the CPU it frees on what the agent waits for; neither changes the film a build delivers. On \bench, the wait falls by \NUM{WaitCutS}{}\% against \rebuild and
\NUM{SeqOursPct}{}\% against \sequential, and the gain holds across models, harnesses and frameworks.

Our scope is 2D video programs whose drawing a runtime can record. Frames of 3D scenes are
rendered exactly but never reused, footage from generative models lies outside the scope, and an edit that changes most
frames gains mostly from scheduling. Exact reuse for 3D and GPU rendering, and agents that plan their edits around what
can be reused, are directions for future research.

\section*{LLM usage statement}
LLM coding agents are the subject of this study: \bench records their sessions, and \cref{sec:exp-agents} runs them
live. GPT-6.1 wrote sixty of \bench's topics, and Claude Opus~5.5 scored the real agents' films blind to the system. A
coding agent helped implement the system and the experiment scripts, ran experiments under the authors' direction and
drafted parts of the text from the authors' notes. The authors checked every result and reference, revised all text and
take full responsibility for the content; no LLM is an author.

\bibliographystyle{ACM-Reference-Format}
\bibliography{main}
\clearpage
\appendix
\section*{Supplementary Material}
\setcounter{section}{0}\setcounter{figure}{0}\setcounter{table}{0}\setcounter{algorithm}{0}\setcounter{property}{0}\setcounter{equation}{0}
\renewcommand{\thesection}{S\arabic{section}}\renewcommand{\thefigure}{S\arabic{figure}}\renewcommand{\thetable}{S\arabic{table}}
\renewcommand{\thealgorithm}{S\arabic{algorithm}}\renewcommand{\theproperty}{S\arabic{property}}\renewcommand{\theequation}{S\arabic{equation}}
\renewcommand{\theHsection}{S\arabic{section}}\renewcommand{\theHfigure}{S\arabic{figure}}\renewcommand{\theHtable}{S\arabic{table}}
\renewcommand{\theHalgorithm}{S\arabic{algorithm}}\renewcommand{\theHproperty}{S\arabic{property}}\renewcommand{\theHequation}{S\arabic{equation}}

\section{\bench in detail}
\label{sec:supp-bench}

A \emph{session} records one agent making one film: every version it committed, every tool call in order with its
timing and outcome, the agent's think time before each call, and the class of edit each build follows.
The \emph{real track} has \NUM{RealSessions}{} sessions of Claude Code~\cite{claudecode} with Claude Opus~5.5~\cite{claudemodels} making
explainer films with cv.js, our Canvas~2D~\cite{canvas2d} runtime (\cref{fig:gallery}). These sessions hold
\NUM{RealVersions}{} versions, \NUM{RealCalls}{} tool calls and \NUM{RealBuilds}{} builds. Their films last
\NUM{FilmPTenS}{} to \NUM{FilmPNinetyS}{}\,s between the 10th and 90th percentiles and are narrated in Chinese and
English. Of the track's builds, \NUM{EditFilesPct}{}\% follow edits that add or remove files, \NUM{EditLinesPct}{}\%
follow edits of narration lines, \NUM{EditZonePct}{}\% follow edits of one scene's script and \NUM{EditSharedPct}{}\%
follow edits of shared scripts or data. The track holds every session recorded over two weeks that built at least two
versions. A scripted user issued every change request. Of these sessions, \NUM{HeldoutSessions}{} are on
\emph{held-out} topics, which GPT-6.1~\cite{openaimodels} wrote for \bench. Each held-out
topic has three rounds of change requests that build on one another.
The \emph{model track} has \NUM{ModelSessions}{} sessions of Claude Sonnet~5.5 and Haiku~5.5~\cite{claudemodels} on \NUM{ModelTopics}{} test topics. The
\emph{Remotion track} brings a second framework, Remotion~\cite{remotion}, with synthetic edits. For each of two films
we wrote and each of three random seeds, a generator applies 24 edits in turn. The generator draws each edit's class and
the think time before each build from those of the real track. This gives \NUM{RxVersions}{} versions in
\NUM{RxSessions}{} sessions. The real track is split by topic. The \NUM{DevSessions}{} sessions on topics used to tune
any parameter form the development split. The other \NUM{TestSessions}{} sessions, on \NUM{TestTopics}{} topics that
include all of GPT-6.1's, form the test split. \Cref{sec:experiments} reports results on the test split. The timed replays take one session from each of \NUM{TimedSessions}{} topics, drawn with a fixed seed. The reuse study takes the \NUM{Pairs}{}
pairs of adjacent delivered versions of \NUM{PairSessions}{} test-split sessions.

\paragraph{Format.}
A session directory holds four entries. \texttt{session.json} gives the track, framework, topic, date, narration
language, film length, the agent's model and harness, the machine profile, the framework's commit, counts, edit classes,
the voice pool and what was scrubbed. \texttt{manifest.jsonl} gives each version's files by content hash.
\texttt{calls.jsonl} gives, per call, the tool, the agent's arguments with the project path abstracted, the version the
call saw, the call's start, recorded duration and exit, a build's verdict time and edit class, and the think time
before the call. \texttt{objects/} holds the file contents by hash. A call's version is the version committed within
2\,s of the call's start or, if there is none, the newest version before the call.

\paragraph{Held-out topics.}
GPT-6.1 wrote the held-out topics of the \NUM{HeldoutSessions}{} sessions in five kinds (concepts, data,
systems, teaching, products) and two languages. Each topic has three rounds of stacked change requests: a local change,
a global change, and a mixed change that takes something back.

Replay is closed-loop~\cite{schroeder2006openclosed}. A system executes a session's calls in order, each call after
the previous one has answered, either at once or after the recorded think time. The system's caches start cold and
persist across versions but not across sessions.
\par\noindent\textbf{Correctness gate.} Approximate reuse is judged by image similarity~\cite{teacache}. \bench instead
requires the answers of a build from scratch before any time counts, which is the \emph{from-scratch consistency} of
incremental computation~\cite{hammer2015adapton}. A build must pass or fail as the build from scratch does, and it must
report no error that the build from scratch does not report. Every captured frame of every delivered film must equal
the frame that a render of that version from scratch captures. The sound must match within the tolerances given below.
\par\noindent\textbf{Metrics.} Renderers are compared by render time~\cite{kamishibai}, which misses what blocks the
agent. The \emph{wait} therefore runs from a build's call until the build answers the agent. A build that fails a check
answers at the verdict, and a build that delivers answers after the film and its checks. We also report the time to
verdict, frames rendered, CPU-seconds and peak memory.

\paragraph{The correctness gate in full.}
The reference for a verdict is a build from scratch. When both a system and this reference ran every check, the two
verdicts must hold the same errors, each given by its check, element and time. Every frame that a system captures for a
delivered film is compared with one reference shared by all systems, the frames of a render from scratch with no cache
and no key. The encoded film must hold every captured frame. It must also decode as the same system's render from an
empty cache on as many cores decodes, since the encoder's output follows the segments that the system cuts and the
cores it has. The sound is compared with that of the render from scratch. It must have the same tracks and channels,
lengths within a frame, and loudness within 0.5\,LU overall and within 1\,dB in its 20-ms envelope.

\paragraph{The checks.}
A build's verdict comes from machine checks of the film and a lint of its sources (\cref{tab:checks}). Most checks
read what a frame draws, its texts with their boxes and its image draws. They sample the film every 40\,ms, and the
overlap check samples it every 0.5\,s. Three checks read the pixels of sampled frames. Two checks read the sound and the
encoded film and therefore need the film. A passing build runs these two after its render (\textsc{FilmCheck} in
\cref{alg:build}). An error stops the build, and a warning is returned with the film. Across versions, a check keeps
its record at a sampled time under the cascade's first two tests, that is, when no changed script reaches that time and
the values read by the frame at that time are unchanged. After a change to any other file, the checks take every record
again.
\begin{table}[h]
\centering\footnotesize
\caption{The checks behind a verdict, by what each reads.}
\label{tab:checks}
\begin{tabular}{@{}lp{0.7\columnwidth}@{}}
\toprule
Check & Reports \\
\midrule
\multicolumn{2}{@{}l}{\emph{A frame's texts and image draws}} \\
Overlap & text over a subject, another text or a subtitle, outside the safe area, or cut by the frame edge \\
Reading & text on screen for less time than reading it takes \\
Timing & text shown more than 0.5\,s before the narration says it \\
Subtitles & a subtitle over a picture without a box \\
Cuts & a picture still moving at a cut, or filling the frame for under 1\,s \\
Resolution & an image drawn larger than its resolution supports \\
Data & bars out of proportion to their labels, numbers absent from the data files \\
\midrule
\multicolumn{2}{@{}l}{\emph{Pixels of sampled frames}} \\
Frames & blank, dark or blown-out frames, every 0.5\,s \\
Style & a scene's frame far from the style's colors \\
Determinism & a scene's frame that differs when rendered again in another order \\
\midrule
\multicolumn{2}{@{}l}{\emph{The sound and the film}} \\
Ducking & music not lowered under the voice \\
File & loudness, true peak, black frames \\
\midrule
\multicolumn{2}{@{}l}{\emph{The sources}} \\
Timeline & cues or subtitles outside the film, locked items changed \\
Lint & the project's conventions for storyboard, assets, style and scenes \\
\bottomrule
\end{tabular}
\end{table}

\paragraph{Baselines in detail.}
Input hash re-renders every frame of a version that changed. Scene windows re-render the frames inside the old and
new windows of every scene whose file changed. When a file that no scene owns changed, they re-render every frame. Read
sets re-render a frame when the files of the parts it draws or the set of parts it draws changed, or when it read a
value that changed. Kamishibai-style fingerprints follow the hand-written tokens of the incremental renderer
kamishibai~\cite{kamishibai}. Typical fingerprints key a part by its file, its own start and end and the local time.
Careful fingerprints add every shared script and the whole timeline. The animation engine Manim~\cite{manim} keys an
animation call by hashing the call's arguments and the scene's objects as they stand. Manim-style per-call caching keys
a scene in the same way, by its file, the shared scripts, its length, every value its frames read and every value its
script read at load, such as a scene's start or a line's timing. It does not key a scene by where the scene sits, and
it keys the subtitles also by the shared scripts that style them. We checked this cache against Manim CE itself on four
kinds of edit. Complete keys cut the film into scenes or into narration lines and reuse a unit whole when the same
sequence of frame keys was rendered before anywhere in the film. They are therefore exact and follow units in time.
They bound what a cache of whole scenes or lines with complete keys could reuse. The similarity rule reuses a frame as
training-free caches reuse features between similar denoising steps~\cite{teacache}. It renders every frame at a
quarter of its size and shrinks the frame to a $32\times18$ gray thumbnail. It reuses the previous version's frame at
the same time when the mean absolute difference of the two thumbnails' gray levels (0--255) is below a threshold, which
we sweep from 0.1 to 8. HyperFrames~\cite{hyperframes}, an HTML video framework for agents, renders ports of the
Remotion track's versions, each port at 1080p with as many frames as Remotion's own render.

\paragraph{What the scrub keeps.}
Every replaced file becomes a placeholder of the same size or length, chosen by the original's content hash. The
same original therefore always becomes the same placeholder, and the scrub keeps which frames change between versions.
A font is kept when its license allows. Otherwise a substitute font holds every character the narration uses. Voices
become tones over the samples where speech was, so the narration's cut points and the timeline are unchanged.
Speech-recognition caches are re-keyed. All numbers in the paper are computed on the scrubbed sessions, which are the
ones released.

\paragraph{Statistics.}
\begin{table}[h]
\centering\footnotesize
\caption{The three tracks of \bench: sessions, versions, tool calls and builds; films, builds and the agent's think time before
a call in seconds, median [10th, 90th percentile]; the share of builds by edit class. Models: Claude Sonnet~5.5 and
Haiku~5.5.}
\label{tab:stats}
\resizebox{\columnwidth}{!}{\begin{tabular}{lrrr}
\toprule
 & Real & Models & Synthetic \\
\midrule
Sessions & 123 & 20 & 6 \\
Versions & 1{,}525 & 215 & 150 \\
Tool calls & 2{,}078 & 289 & 150 \\
Builds & 1{,}314 & 188 & 150 \\
Calls per session & 16.9 & 14.4 & 25.0 \\
Chinese / English & 69 / 54 & 16 / 4 & 0 / 6 \\
Film (s) & 80.8 [52.0, 145.8] & 79.1 [56.7, 177.8] & 31.9 [24.5, 40.4] \\
Build (s) & 25.5 [3.1, 63.6] & 14.8 [2.1, 49.6] & -- \\
Think time (s) & 8.0 [0.0, 87.5] & 15.3 [0.0, 119.2] & 12.8 [4.7, 105.3] \\
Builds per session & 10 [5, 16] & -- & -- \\
Rounds per session & 4 [3, 7] & -- & -- \\
Model cost per session (\$) & 2.85 [1.73, 6.45] & -- & -- \\
\midrule
Edit class: new or removed files & 39.0\% & 41.1\% & 33.3\% \\
Edit class: narration lines & 26.1\% & 23.2\% & 27.1\% \\
Edit class: one scene's script & 12.6\% & 16.1\% & 20.1\% \\
Edit class: shared scripts or data & 20.4\% & 18.4\% & 15.2\% \\
Edit class: a style script & 0.3\% & 0.6\% & 0.0\% \\
Edit class: no change & 1.6\% & 0.6\% & 4.2\% \\
\bottomrule
\end{tabular}
}
\end{table}
The Remotion track's edit classes differ from the real track's by \NUM{StatsEditTvd}{} in total variation, and its
think times differ from the real track's by \NUM{StatsGapKs}{} in the Kolmogorov--Smirnov statistic (\cref{tab:stats}). \Cref{tab:rx-edit} gives the Remotion track's wait by how much of a film a build changes.

\paragraph{Rules.}
Replays are closed-loop. Caches start cold and are never shared across sessions or replays. A system's timings count only if every call passes the gate. We built the system and set its parameters on the development split. The early render's
control group and rule, the fast checks, the page's part of the key and the place of the key evaluation were designed
with the test split's replays in view. The data come with a key trace, which gives every frame's key and read values
for each rendered version, renumbered within the session. They also come with an OpenTelemetry export, with one
\texttt{invoke\_agent} span per session and one \texttt{execute\_tool} span per call. A datasheet and Croissant metadata
describe the data.

\section{Why the keys are sound, and what is audited}
\label{sec:supp-sound}
\paragraph{The guarantee.}
On one renderer configuration (runtime, browser, rasterizer, frame size), if a frame's key equals a key rendered before,
its pixels equal the pixels rendered under that key, provided that (i) the frame's drawing is deterministic, (ii) the key holds the frame's
structural inputs (the page, the set of registered parts and their order),
(iii) every part used only the supported interfaces below, and (iv) the rasterizer draws the normalized commands as the
normalization rules below characterize it. This is Property~\ref{prop:sound} with each of its provisos spelled out.
A part that uses an interface outside the table below is unkeyed for that frame and always rendered.

\paragraph{How each interface enters a key (Canvas 2D).}
Each part draws its share of the frame at time $t$ in its function \texttt{render(t)}.
\begin{center}\scriptsize
\begin{tabular}{p{0.42\linewidth}p{0.5\linewidth}}
\toprule
Interface & Key \\
\midrule
Drawing calls (\texttt{fillRect}, \texttt{fill}, \texttt{stroke}, \texttt{clip}, \texttt{fillText}, \texttt{drawImage}, \texttt{putImageData}, \dots) & method, arguments, and a snapshot of the state at the call \\
State (\texttt{save}/\texttt{restore}, transforms, alpha, compositing, filter, font, text alignment, line styles, dashes, shadows, smoothing) & a state stack kept in call order, snapshot at each drawing call \\
Paths and \texttt{Path2D} & the command sequence with its transforms, keyed when filled, stroked or clipped \\
Gradients and patterns & construction arguments and every color stop up to use; pattern source by content and its transform \\
Images, fonts, data files & content hash of the asset \\
Off-screen canvases as sources & the hash of their command sequence since creation or the last whole overwrite \\
\texttt{getImageData}, \texttt{putImageData} & the written bytes \\
Commands that set state but never draw & dropped at the part's end (no pixels) \\
\midrule
Unregistered image URLs, video sources, WebGL canvases as sources, off-screen canvases changed outside \texttt{render(t)}, resizing a canvas mid-frame & \textbf{not keyed: the part is rendered} \\
DOM outside the canvas, WebGL / three.js, post-processing & \textbf{global inputs} \\
\bottomrule
\end{tabular}
\end{center}

\paragraph{Determinism.}
Reads of \texttt{Math.random}, \texttt{Date.now} and \texttt{performance.now} inside \texttt{render(t)} are recorded. The
checks render one frame of every scene twice, in different orders, with the unmodified runtime. They compare the two
renders' keys, or their full-size pixels for unkeyed frames. The checks report the recorded reads as defects.

\paragraph{Normalization rules.}
The rasterizer draws text from its glyph atlas at a baseline snapped to whole device pixels, so such text is keyed
by the snapped baseline. The drawing itself is unchanged, and the key follows the rasterizer. This snapping applies to
text drawn by \texttt{fillText} without a width limit, under an axis-aligned transform with positive scales and no
filter or shadow, at a glyph size of at most 256 device pixels, the atlas's limit, and with a device baseline at least
0.02 pixels from a rounding boundary. Any other text enters a key with its coordinates as given. In a sweep over fonts,
sizes, equal and unequal scales, all baselines, fill and stroke, color, alpha, spacing and the atlas edge, every such
key stands for one picture.
An off-screen canvas or layer forgets its pixel history when it is overwritten whole from a clean state, by a
whole-canvas \texttt{clearRect} or \texttt{putImageData} or by an opaque \texttt{fillRect} under the identity
transform. The canvas's hash then restarts with its drawing state. Any other drawing keeps the history in the hash.
Chromium keeps a raster state per 2D canvas that follows what the canvas drew before, and resizing the canvas to zero
and back clears this state. Every canvas (the main one, the caption layer and every other layer) is therefore resized
to zero and back before each captured frame. Three renders of a version then agree frame for frame. A frame evaluated
only for its key is drawn on contexts reset in place. It is drawn in full when its drawing reads pixels back or when
the page holds a context other than 2D. The 2D canvas is rasterized in software at full scale. On Remotion pages, what
a frame draws is the laid-out page, with its elements under their computed styles and its resources by content. An
element that cannot be visible, with opacity zero, is keyed by its box alone. Text on these pages is drawn in
grayscale, since with subpixel text a frame's pixels depend on what the page drew before.

\paragraph{The cascade in full.}
\label{sec:supp-cascade}
\Cref{alg:reuse} gives the keys and the film of a version. The algorithm's levels are the stages of \cref{fig:arch}.
Static impact is level 1, values read is level 2, the frame key is level 3 and the render is level 4.
Let $T_P$ be the frame times of $P$. Let $Q$ be the last \emph{key pass} in the same configuration, the last version
whose keys were computed, with key $k^Q_t$ and read set $R_t$ for each frame $t$. Let $D=P\ominus Q$ be the files that
changed. With no $Q$, every key is evaluated. A scene's script draws only inside the scene's time window. Let $\Omega$
be the union of the windows $\omega$, widened by one frame, that the parts reached by the changed scripts have in $Q$
and in $P$ (\cref{alg:l1}). A frame outside $\Omega$ therefore keeps its key if no value it read changed
(\cref{alg:l2}). While rendering, the runtime records through a read probe the values $R_t$ that each frame reads, such
as the timeline, the style, a part's configuration or an asset entry. The runtime then compares their hashes
$\eta_P(R_t)$ and $\eta_Q(R_t)$, so editing one entry of the style invalidates only the frames that read that entry.
Values read at load time are recorded per script rather than per frame. A script whose load-time values changed starts
the reach as a changed script does, so the windows of the parts that this script reaches join $\Omega$. A change that
no part owns, such as a data file read whole, the page or the runtime, reaches every frame. The film's length counts as
a value read by the frames that use it. Level 2 cannot apply when $P$ loads other resources than $Q$ or when a value
read at load has no known reader. Then no key is kept, and every frame goes to level 3. Let $K$ be the frames that keep
their keys. Before these keys are used, up to four frames $U(K)$ drawn at random from $K$ are evaluated afresh. A fresh
key that differs from the kept one sends every frame to level 3 (\cref{alg:audit}). \Cref{prop:kept} states when a kept
key is the key that level 3 would compute. The audit tests this property on the film at hand. In the runs we report,
\NUM{KeptPasses}{} key passes kept \NUM{KeptByValue}{} keys, and each kept key passed both tests.

\begin{algorithm}[t]
\caption{Keys and film of a version $P$ (symbols in \cref{sec:supp-cascade}).}
\label{alg:reuse}
\small
\begin{algorithmic}[1]
\Require $P$; last key pass $Q$ with keys $k^Q$, read sets $R$; stores of frames $\mathcal C$, segments $\mathcal S$, films $\mathcal F$
\Ensure film $F_P$
\State \textbf{if} $P\in\mathcal F$ \textbf{then return} $\mathcal F[P]$ \label{alg:film}
\State $\Omega\gets\omega_Q(D)\cup\omega_P(D)$, $D=P\ominus Q$ \Comment{level 1} \label{alg:l1}
\State $K\gets\{t\notin\Omega: k^Q_t\neq\bot,\ \eta_P(R_t)=\eta_Q(R_t)\}$ \Comment{level 2} \label{alg:l2}
\State \textbf{if} $\exists\,t\in U(K):\kappa_P(t)\neq k^Q_t$ \textbf{then} $K\gets\emptyset$ \label{alg:audit}
\State $k_t\gets k^Q_t$ \textbf{if} $t\in K$ \textbf{else} $\kappa_P(t)$, $t\in T_P$ \Comment{level 3} \label{alg:l3}
\For{$s\in\mathrm{Seg}(P)$} \Comment{level 4} \label{alg:seg}
  \State \textbf{if} $k_s\in\mathcal S$ \textbf{then} $F_s\gets\mathcal S[k_s]$; \textbf{continue}
  \State $I_t\gets\mathcal C[k_t]$ \textbf{if} $k_t\in\mathcal C$ \textbf{else} $\rho_P(t)$, kept in $\mathcal C$ if $k_t\neq\bot$; $t\in s$ \label{alg:l4}
  \State $F_s\gets\mathrm{enc}(I_s)$, kept in $\mathcal S$ if $\bot\notin k_s$ \label{alg:enc}
\EndFor
\State \Return $\mathcal F[P]\gets F_{s_1}\Vert\cdots\Vert F_{s_m}$
\end{algorithmic}
\end{algorithm}

The remaining frames get the key $\kappa_P(t)$ of the commands they issue (\cref{alg:l3}), evaluated without
capturing the frame. A frame that draws through an interface outside the supported set gets no key, written $\bot$.
Every part that draws the frame, such as a scene, an overlay or a caption, records the drawing commands it issues. Each
command is recorded as its method, its arguments and the drawing state, which holds the transform, styles, font, clip
and compositing. Images and fonts enter the key by their bytes in the version's immutable store. Gradients and
off-screen canvases enter by their content, and the page around the canvas enters by its elements and styles.
The key hashes the parts in drawing order. Outputs are stored per renderer configuration.
On a recorded film of \NUM{PerfFrames}{} frames, the key pass costs \NUM{KeyCpuMs}{}\,ms of CPU per frame, against
\NUM{ShotCoreMs}{}\,ms for a full render with sound and encoding. Levels 1--2 skip even the key evaluation for the
frames they keep. In the runs of \cref{sec:experiments}, the key pass takes a median of \NUM{KeyFilmMedM}{}\,s per film
on 8 cores and \NUM{KeyFilmMedS}{}\,s on 32. That is \NUM{KeyRenderPctM}{}\% of a render's time on 8 cores and
\NUM{KeyRenderPctS}{}\% on 32, and \NUM{KeyWaitPctM}{}\% of a film's wait on 8 cores and \NUM{KeyWaitPctS}{}\% on 32.
\begin{property}[Kept keys]
\label{prop:kept}
Let $Q$, on the configuration and frame rate of $P$, hold $k^Q_t=\kappa_Q(t)$. If $t\notin\Omega$ and
$\eta_P(R_t)=\eta_Q(R_t)$, then $\kappa_P(t)=k^Q_t$, provided the assumptions of Property~\ref{prop:sound} hold, each
part is attributed to the script that registers it, a script affects frames outside its parts' windows (an overlay's
window is the film) only through the names it declares, and the probed data (the timeline, the style, asset entries)
reach a part only through reads the probe records: the part's own while drawing and its scripts' at load.
\end{property}
\paragraph{Why a kept key is the level-3 key.}
Every key of $Q$ was evaluated or kept under \cref{prop:kept}, so by induction $k^Q_t=\kappa_Q(t)$. On one
configuration, $\kappa(t)$ hashes the frame's size, background, fonts, page and camera, then each drawing part's name,
z-order and commands in order. Each part starts from a reset drawing state, so by Property~\ref{prop:sound}'s
determinism its commands are a function of its code, window, $t$, load-time values, reads while drawing and assets. A
scene draws only inside its window. $\Omega$ holds the windows, in $Q$ and $P$, of the parts of every script reached
through declared names from a changed script. A script counts as changed when its code or its load-time reads changed,
and a data script when its values changed. $\Omega$ holds every frame if a reached script registers an overlay or
writes into shared state. By the third proviso, no changed script affects a frame $t\notin\Omega$, so the same parts
draw in the same order, with the same code, windows, load-time values, fonts, page and camera. By the fourth proviso,
data reach these parts only through those load-time values and through $R_t$, which holds their reads while drawing,
the frame size and the background, and $\eta_P(R_t)=\eta_Q(R_t)$. The page, style sheets and font files belong to the
configuration. A changed asset file either comes with changed entries that its readers read or sends every frame to
level~3. Equal inputs give equal commands, so $\kappa_P(t)=k^Q_t$. Read sets (\cref{tab:exact}) tie a frame to its
reads while drawing and to its scene's file. They miss a change through code that the frame calls in another scene's
file, and they miss a changed value read at load, such as a scene's start, so they reuse stale frames. Of the
\NUM{ReadSetStale}{} stale frames they reuse on the adjacent versions of \cref{tab:exact}, \NUM{ReadSetOmegaPct}{}\% lie
in the windows that the changed scripts reach. The other \NUM{ReadSetLoadPct}{}\% lie in the windows reached from a
script whose values read at load changed. All of them lie in $\Omega$, so levels 1--2 refuse every one.
\Cref{tab:kept} lists what enforces each condition. Over \NUM{AuditKeptFrames}{} frames of \NUM{AuditKeptPairs}{}
pairs, the key audit finds every kept key equal to its fresh evaluation (\cref{sec:exp-exact}).
\begin{table}[t]
\centering
\caption{What enforces each condition of \cref{prop:kept}; ``level 3'' sends every frame to be evaluated.}
\label{tab:kept}
\scriptsize
\begin{tabular}{@{}p{0.27\linewidth}p{0.67\linewidth}@{}}
\toprule
Condition & How it is enforced \\
\midrule
$k^Q_t=\kappa_Q(t)$ & $\bot$ is never kept; keys are taken after the loads a draw asked for; a kept key carries its read
set forward, and every kept key passed both levels (induction over key passes) \\
Configuration, frame grid & one index per runtime, browser, fonts, page, size, scale and the renderer's frame rate; a
size or background set by a script is read by every frame \\
Same parts, names, order & $\Omega$ holds, in $Q$ and $P$ ($\pm1$ frame), the windows of every part a reached script
registers; an overlay reached, two parts of one name, or a script added or removed: level 3 \\
Code & reach follows the names a script declares or binds (declarations at any depth, implicit globals, members of the
global object under any of its names, string property names); a computed global access reaches every name; a write into
shared state or the page: level 3 \\
Values read at load & recorded per script, cues included; a script whose values changed starts the reach \\
Values read while drawing & the timeline, the style, cues, asset entries, sound durations, the film's length, the
registration, size and background are probed per frame, raw reads and property queries included; an opaque value or a
write while drawing: no key \\
Start state & every canvas reset before each frame, each part drawn between a save and a restore, an inherited path or
an off-screen canvas drawn twice: no key; \texttt{Math.random}, \texttt{Date.now}, \texttt{performance.now} while
drawing: no key \\
Fonts, page, camera & faces by their bytes, the page by its elements and styles after each frame, the camera's value:
all in the key; their files in the index \\
Assets & by the hash of the bytes decoded; a file changed without its registry entry: level 3 \\
Changes no part owns & the page, style sheets, fonts, runtime and browser in the index; a data file read whole: level 3 \\
\bottomrule
\end{tabular}
\end{table}

Level 4 renders the film (lines~\ref{alg:seg}--\ref{alg:enc}). The film is cut into segments $\mathrm{Seg}(P)$ at
scene starts and every second after each scene start. A segment $s$ whose sequence of keys $k_s$ was encoded before is
copied from the segment store $\mathcal S$. The other segments are encoded ($\mathrm{enc}$) from their frames $I_t$,
each taken from the frame store $\mathcal C$ by its key or rasterized ($\rho_P$). Unkeyed frames and the segments
holding them are never stored. A version built before returns its film whole from the film store $\mathcal F$
(\cref{alg:film}). Segments hold pictures only. The narration is voiced again for the lines that changed, and the sound
is mixed for every version and muxed with the segments. A segment whose keys are unchanged is therefore copied under a
new narration. Work ahead runs the same render without sound and writes no film.
\begin{property}[Fewest captures]
\label{prop:fewest}
Given the outputs stored so far, none evicted, \cref{alg:reuse} captures one frame for each distinct key of $P$ that has
no stored output, and every unkeyed frame; no cache holding the same outputs that reuses a frame only under an equal key
captures fewer.
\end{property}
A frame whose key is already being captured waits for that capture, so each missing key is captured once.

\par\noindent\textbf{What is reused.}
A \emph{frame store} holds the rendered frames by key, per renderer configuration. A build takes from the store every
frame whose key the store holds and renders the rest, each missing key once. The film is cut into segments at scene
starts and every second after each scene start, so a moved scene keeps its segments. A segment whose keys were encoded
before is copied, and a version built before returns its film whole. The checks reuse their results through the
cascade's first two tests.

\section{Scheduling details}
\label{sec:supp-sched}
\paragraph{Mechanisms and when each is used.}
The checks' text sweep samples the film every 40\,ms, and their frame checks sample it every 0.5\,s. Within a build,
the sweep runs over $k$ browser pages in parallel (par), with $k=8$ on 32 cores or more and $k=\max(2,\min(4,c-4))$ on
$c$ cores below 32. The parallel sweep is used on $c\ge6$ cores without a software-only rasterizer. The source lint runs
alongside the checks (dag). The fast checks (vp) draw a sampled frame on reset contexts, as the evaluation of a key
does, instead of on a new canvas. They draw a frame whose code reads pixels back in full. The fast checks also keep the
source map, which maps drawn objects to source lines, off the verdict's path. A check without errors returns at once,
and a check with errors scans the frames its errors name for their source lines. After the verdict, a separate process
at the lowest priority rebuilds the map. The early render (opt) starts alongside the checks on pages of its own, in a
control group of its own marked idle. A passing verdict gives the group a normal share and the first claim on memory. A
failing verdict kills the group whole. A failure known from the files before the verdict, such as a stale narration
text or a lint error, stops the early render at once. The early render starts when $\hat p>\theta_c$
(\cref{eq:early}), with $\hat p=(a+4m)/(n+4)$. Here $n$ counts the session's last five builds of the same kind of edit,
narration changed or not, or the last five of any kind while that kind has fewer than two. The count $a$ holds those
among them whose checks passed. The rate $m$ is the development split's pass rate, $1/33$ before a session's first build
and $186/360$ after it. The threshold $\theta_c$ is interpolated in $\log_2 c$ between the machines of
\cref{tab:interference} and held at the ends of that range. A session's first build thus never starts the render early.
A later build always does on 4, 32 and 64 cores, and on 8 cores unless the five builds counted all failed. The capture
workers evaluate the keys, each worker for its own segments, instead of a pass before the capture (kw). On machines
below 16 cores, builds fail fast (ff). The data, style, determinism and timeline checks, which seldom fail, then run
only once the other checks found no error that stops the build. Operations run at the priorities of their consumers
(pri), which are normal priority, the idle group and nice 19. Within a render, the audio chain runs alongside the frames
(aud), capture pages open early (pre), each worker encodes a segment while it shoots the next (enc), and the workers
are sized to the machine (wk). \rebuild uses every mechanism that applies on the machine except par, dag, vp, opt, kw,
ff and pri. It reuses only a whole film whose version did not change. A shared execution service with warm pages serves
its callers in turn. Renders get no fixed share of the cores.

\par\noindent\textbf{Within a build: the verdict first.}
On half of the builds the agent waits only for the verdict, whose checks sweep the whole film. \sys runs the checks
first and in parallel. Their sweep is cut by time over several browser pages, and the lint runs alongside the sweep.
The \emph{fast checks} of \sys also leave out what a verdict does not need. The checks that read what a frame draws,
not its pixels, draw their samples as the evaluation of a key does (\cref{sec:reuse}). The map from drawn objects to
source lines is built after the verdict. Where cores are scarce and the checks that seldom fail would delay the
verdict, a build also fails fast, as in test prioritization~\cite{rothermel2001apfd,prado2020ntr}. The checks that
seldom fail then run only once the others found no error, and the build answers at the first errors that stop it.

\par\noindent\textbf{Within a build: the render as a bet.}
Starting the render alongside the checks is a bet on the outcome, like the sandboxes that agent serving
prewarms~\cite{specbox}. The early render saves render time when the version passes and slows the checks in either
case. At idle priority, the render slows the checks by a fraction $d$ through the memory, kernel and I/O they share,
and it keeps a share $s$ of its own speed. The bet therefore pays when the version passes with probability above
$\theta_c=d/\bigl(s(1+d)\bigr)$ (\cref{sec:supp-vfirst}). Idle priority keeps the bet cheap. The measured $\theta_c$ is
\NUM{IdleNeedM}{} on 8 cores and \NUM{IdleNeedS}{} on 32. \Cref{alg:early} takes the bet when $\hat p>\theta_c$, where
$\hat p$ estimates that probability from the session's last builds. A session's first build, which rarely passes, thus
waits for its verdict, and later builds render early (\cref{fig:teaser}c). The render starts early in
\NUM{EarlyShareS}{}\% of all builds on 32 cores and in \NUM{EarlyShareM}{}\% on 8.

\paragraph{The verdict first, argued.}
\label{sec:supp-vfirst}
Starting the render alongside the checks delays the verdict by $\delta$ whatever the outcome. The version passes with
probability $p$, and when it passes, the early start saves $g$ of the render's own time. The expected waits with the
render started early and after the verdict therefore differ by
\begin{equation}
  \mathbb E[W_P^{\mathrm{early}}]-\mathbb E[W_P^{\mathrm{after}}] = \delta - p\,g ,
  \label{eq:early}
\end{equation}
and starting early pays when $p\,g>\delta$. The following property holds in this model.
\begin{property}[Verdict first]
\label{prop:vfirst}
On $c$ cores where operations preempt at no cost, fit in memory and interfere only by taking cores, running the
verdict's operations at strict priority finishes the verdict when the same order of those operations finishes it with
nothing else running, and starting the film's operations on the cores it leaves idle, instead of after it, delays
neither the verdict nor the film, whose frames are captured independently; then $\delta=0$, and starting early never
lengthens the wait.
\end{property}
The verdict's operations get every core they can use whatever else runs, so they run as they would alone. Film work
done before the verdict only shortens what is left of the film after the verdict, since no frame waits for another. By
the same argument, work ahead in the think time costs the next build nothing in this model, and it saves the next
build every rendered frame that the next version draws.
On a real machine, the checks also share memory, kernel and I/O with a capture running alongside them. While both run,
the checks advance at $1/(1+d)$ of their speed alone, and the render advances at a share $s$ of its own speed. Each
second of overlap therefore delays the verdict by $d/(1+d)$ seconds and saves $s$ seconds of the render, however long
either runs. Hence $\delta/g=d/(s(1+d))$, and by \cref{eq:early} starting early pays when $p>\theta_c=d/(s(1+d))$.
With the render in the idle group, $d=\NUM{IdleDM}{}$ and $s=\NUM{IdleKeptM}{}$ give $\theta_c=\NUM{IdleNeedM}{}$ on
8 cores, and $d=\NUM{IdleDS}{}$ and $s=\NUM{IdleKeptS}{}$ give $\theta_c=\NUM{IdleNeedS}{}$ on 32
(\cref{tab:interference}).
After a pass, a render started early runs at normal priority and first in the memory queue. The checks that need the
film (\textsc{FilmCheck}) run after the render.
\paragraph{Across agents sharing a machine.}\label{sec:colocate}
Under the capacity constraint, the builds of all agents on a machine draw cores and memory from one ledger sized from
measured footprints. Agent serving similarly shares servers among agents~\cite{justitia} and meters resources per
tool call~\cite{agentcgroup}. A render's workers follow what the render is granted, and work ahead yields when blocking
work cannot fit. The ledger's entries follow the memory admission below.

\paragraph{Memory admission.}
A render asks a ledger, kept per memory budget, for the memory of its plan. This memory is estimated from
measurements on this machine: 0.4\,GB plus 0.9\,GB per capture worker at 1080p, 5\,MB per second of film for the audio
chain alongside the frames, and 0.15\,GB per warm page held by the execution service. The workers are first capped by
the cores the render may use and then sized by the memory granted. The audio runs alongside the frames only if it fits
in the grant. When the grant falls short of the plan, the render first drops the audio's overlap and then workers.

\paragraph{Pages without the GPU.}
Chromium gives every browser a GPU process. With the NVIDIA driver, these processes wait on a lock of the driver that
all of them share, whatever cores or GPU they use. On a \NUM{NgFilmSec}{}\,s film at 1080p, three renders at once each
took \NUM{NgSlowGpu}{}$\times$ as long as one render alone. A film whose code never asks for WebGL is shown with the GPU
off (\texttt{--disable-gpu --disable-software-rasterizer}), so no process opens the driver. Its Canvas 2D is rasterized
on the CPU at full size either way, so the frames are the same (\NUM{NgFramesSame}{} frames of \NUM{NgFilms}{} films).
No render then waits on the lock. A render takes \NUM{NgRenderNone}{}\,s instead of \NUM{NgRenderGpu}{}\,s, and three
renders at once each take \NUM{NgSlowNone}{}$\times$ as long as one. Every system in \cref{sec:experiments} runs this
way, and the mode is part of every cache of pixels. A film counts as one that may use WebGL when its code mentions
WebGL anywhere. Such a film is drawn by SwiftShader, which runs the GPU interfaces on the CPU, with the Vulkan loader
shown only SwiftShader's driver, so no process opens the graphics device either. Its 2D canvases are rasterized on the
CPU as in every other film, since on SwiftShader's GPU path a canvas's pixels depend on what the page drew before. Two
renders from scratch of one version then agree frame for frame. WebGL frames are never keyed and are always rendered
afresh. On the adjacent versions of a test session that draws with WebGL, every frame equals a render from scratch.

\paragraph{Interference.}
A capture running alongside the checks shares the cores and the node's memory, kernel and I/O with them.
\Cref{tab:interference} gives what this sharing costs with the capture in the idle control group, on machines of 4 to
64 cores. It compares the time of the checks and lint of one build of a development film with their time alone, and the
capture's frame rate with its rate alone. Each value is the mean of five repetitions. With the capture niced to 10 inside the build's own group instead, $d$ is \NUM{NiceCallDM}{} on 8 cores and $\theta_c$ rises to \NUM{NiceCallNeedM}{}.
\begin{table}[h]
\centering\footnotesize\setlength{\tabcolsep}{6pt}
\caption{The early render alongside the checks: the checks' slowdown $d$, the share $s$ of its rate the render keeps and
the threshold $\theta_c$ of \cref{eq:early}.}
\label{tab:interference}
\begin{tabular}{lrrr}
\toprule
Cores & $d$ & $s$ & $\theta_c$ \\
\midrule
4 & \NUM{IdleDW}{} & \NUM{IdleKeptW}{} & \NUM{IdleNeedW}{} \\
8 & \NUM{IdleDM}{} & \NUM{IdleKeptM}{} & \NUM{IdleNeedM}{} \\
16 & \NUM{IdleDMSixteen}{} & \NUM{IdleKeptMSixteen}{} & \NUM{IdleNeedMSixteen}{} \\
32 & \NUM{IdleDS}{} & \NUM{IdleKeptS}{} & \NUM{IdleNeedS}{} \\
64 & \NUM{IdleDSSixtyFour}{} & \NUM{IdleKeptSSixtyFour}{} & \NUM{IdleNeedSSixtyFour}{} \\
\bottomrule
\end{tabular}
\end{table}

\paragraph{Co-location conditions.}
All conditions run inside one scope that sets the machine's cores, a CPU quota and a memory cap without swap. The
scope's memory events, its kills, are kept. Uncoordinated execution has no priority classes and no ledger, and it runs
one executor per session. Niced renders add the priority classes of \cref{sec:verdict} and nothing else. One control
group per call runs like uncoordinated execution but puts every call in a control group of its own at equal weight.
Each session's executor gets a control group of its own for its daemons. \sys has one ledger of the cores and memory,
the priority classes, and one executor for all sessions, which serves its callers in turn.
Sessions arrive at fixed offsets. Each session replays its calls closed-loop with a quarter of its recorded think time.
Each run takes eight cores of the upper hyperthread half of one socket, whose siblings our jobs leave unused.

\section{Registration and the machine}
\label{sec:supp-versions}

\par\noindent\textbf{Frameworks and hardware.}
Every system renders the same programs at 1080p and 30 frames per second. These are cv.js programs, written for our
Canvas~2D runtime and run in headless Chromium, and Remotion~\cite{remotion} programs, written in React. The machines
are 32, 8 or 4 of one server's cores, with 2\,GB of memory per core. 32 cores are abundant, since a render starts fewer
workers than there are cores. 8 cores, a laptop's size, are constrained, since a render can use more workers than
cores. 4 cores are saturated, since frame capture alone fills them. Each session runs in its own control group on an
AMD EPYC 9755 machine. Every system renders on the CPU, for the reasons given under Pages without the GPU
(\cref{sec:supp-sched}).

\paragraph{Registration.}
We registered the protocol before any formal run. It fixes the split, the machines, the gate and every test. The
tests on 8 and 32 cores form the primary family under Holm's procedure~\cite{holm1979}.
The registered decisions are these. Exactness holds if every pair, at least 1{,}800 random edits and the 32 narration
edits are exported and every frame of their incremental renders equals a render from scratch. For 8 and 32 cores, the
upper end of each ratio's interval must lie below one, one-sided at 2.5\% for the family under Holm's correction. No
resample's ratio reaches one. The largest of the \NUM{HolmResamples}{} resampled ratios is \NUM{HolmMaxRatioS}{} on 32
cores and \NUM{HolmMaxRatioM}{} on 8. For real agents, the measure is the geometric mean over pairs of the ratio of
within-session median waits per build. It is taken on the 14 pairs whose topics development never touched, with all 26
pairs as a secondary analysis. The films' quality is reported as paired differences. Per build, the agent's wait under
\sys relative to its wait under \rebuild on the same build has a median of \NUM{WaitMedS}{}$\times$ on 32 cores (95\%
CI \NUM{WaitMedSCI}{}) and \NUM{WaitMedM}{}$\times$ on 8 (CI \NUM{WaitMedMCI}{}). \Cref{tab:latency} gives the ratios of
total waits.

\paragraph{The machine.}
The machine has 512 hardware threads on two sockets. Timed replays ran in lanes of disjoint cores on the upper
hyperthread half of one socket, whose siblings our jobs left unused. Both systems of a session ran back to back in one
lane. The scaling runs of \cref{fig:scaling} ran one replay at a time.

\section{More results}
\label{sec:supp-results}

\subsection{Comparison with existing systems}
\paragraph{The size of an edit.}
On \bench's Remotion track, the gain follows how much of a film a build changes (\cref{tab:rx-edit}). We group the
builds of three repeats by the share of the film's frames that \sys renders. The agent waits \NUM{RxNoneB}{}$\times$
as long under \sys as under Remotion's renderer when \sys renders no frame, \NUM{RxUnderFiveB}{}$\times$ when \sys
renders under 5\%, \NUM{RxMidB}{}$\times$ when it renders 5 to 20\% and \NUM{RxLargeB}{}$\times$ when it renders 20\%
or more (CI \NUM{RxLargeBCI}{}, resampling the sessions).
\begin{table}[h]
\centering\footnotesize
\caption{The Remotion track's wait under \sys relative to Remotion's renderer, by the share of the film's frames a build
renders; builds count the three repeats, and intervals resample sessions, each with its three repeats.}
\label{tab:rx-edit}
\begin{tabular}{lrrr}
\toprule
Frames rendered & Builds & Sessions & Wait \\
\midrule
None & 162 & 6 & 0.35 \\
Under 5\% & 162 & 6 & 0.46 \\
5--20\% & 78 & 6 & 0.56 \\
20\% or more & 48 & 6 & 0.92 \\
\bottomrule
\end{tabular}

\end{table}
\paragraph{An agent's Remotion sessions.}
The gain on Remotion holds for an agent's own edits. Claude Code with Claude Opus~5.5 made \NUM{RxAgentSessions}{}
films in Remotion, each through three rounds of stacked change requests. These sessions hold \NUM{RxAgentVersions}{}
versions, \NUM{RxAgentBuilds}{} builds and \NUM{RxAgentFilms}{} delivered films. Replayed three times on 8 cores, the
agent waits \NUM{RxAgentWait}{}$\times$ as long under \sys as under Remotion's renderer (CI \NUM{RxAgentWaitCI}{},
resampling sessions, each with its three repeats). On builds that deliver a film, the agent waits
\NUM{RxAgentFilm}{}$\times$ as long. \sys uses \NUM{RxAgentCpu}{}$\times$ the CPU of Remotion's renderer and renders
\NUM{RxAgentFramesPct}{}\% of the frames. Of the \NUM{RxAgentExactOf}{} films the three replays deliver,
\NUM{RxAgentExactFilms}{} equal a render from scratch frame for frame. The others belong to one session, whose film
Remotion's renderer draws differently from run to run.

\subsection{Wait across machines and agents}
\paragraph{Other models' sessions.}
The recorded sessions of the other models differ from Claude Opus~5.5's in how often a build fails and in how much of
a film a version changes. These are the quantities on which the two parts act. On the same ten test topics, a check
stops \NUM{FailSonnet}{}\% of Sonnet's builds, \NUM{FailHaiku}{}\% of Haiku's builds and \NUM{FailOpus}{}\% of Opus's
builds. The median share of a later film's frames that no earlier version drew is \NUM{NewSonnet}{}\% for Sonnet,
\NUM{NewHaiku}{}\% for Haiku and \NUM{NewOpus}{}\% for Opus. The ratios of \cref{tab:latency} follow that mix. They are
\NUM{OtherWaitLo}{}--\NUM{OtherWaitHi}{}$\times$ for the sessions of other models and harnesses and
\NUM{OpusNewWait}{}$\times$ for Opus~5.5 on the $\dagger$ topics, nine topics on which each agent ran its first
session. The parts' own ratios on failing builds and on builds that deliver a film follow the mix as well (the last
block of \cref{tab:ablation}). In Opus~5.5's $\dagger$ sessions, which render a third of the frames, the agent waits
least (\cref{fig:law}). Six of the $\dagger$ topics are development topics.
\Cref{tab:latency-ci} gives the wait under \sys by machine and by agent, with its intervals ($\dagger$: each agent's
first session on nine topics).

\begin{table*}[t]
\centering\footnotesize
\caption{The first two blocks of \cref{tab:latency} with 95\% intervals (2{,}000 resamples of topics), topics and builds.}
\label{tab:latency-ci}
\begin{tabular}{lrrrrrr}
\toprule
Setting & Topics & Builds & Wait & Failed & Film & Frames \\
\midrule
32 cores & 70 & 702 & 0.64 [0.61, 0.66] & 0.39 [0.36, 0.42] & 0.71 [0.68, 0.74] & 0.65 \\
32 cores, think time & 70 & 702 & 0.59 [0.58, 0.61] & 0.41 [0.38, 0.44] & 0.65 [0.63, 0.67] & 0.49 \\
8 cores & 70 & 702 & 0.67 [0.64, 0.69] & 0.43 [0.40, 0.45] & 0.71 [0.68, 0.74] & 0.66 \\
8 cores, think time & 70 & 702 & 0.60 [0.58, 0.63] & 0.43 [0.40, 0.46] & 0.63 [0.61, 0.66] & 0.53 \\
4 cores & 20 & 197 & 0.75 [0.70, 0.80] & 0.46 [0.42, 0.50] & 0.79 [0.72, 0.85] & 0.63 \\
32 cores, Sonnet 5.5 & 10 & 81 & 0.62 [0.55, 0.73] & 0.47 [0.40, 0.57] & 0.66 [0.59, 0.79] & 0.62 \\
32 cores, Haiku 5.5 & 10 & 107 & 0.54 [0.51, 0.60] & 0.46 [0.39, 0.58] & 0.57 [0.54, 0.61] & 0.63 \\
32 cores, DeepSeek V4$^\dagger$ & 9 & 95 & 0.56 [0.50, 0.62] & 0.35 [0.29, 0.42] & 0.67 [0.60, 0.74] & 0.69 \\
32 cores, Codex CLI$^\dagger$ & 9 & 46 & 0.54 [0.37, 0.77] & 0.27 [0.20, 0.60] & 0.63 [0.48, 0.80] & 0.83 \\
32 cores, Opus 5.5$^\dagger$ & 9 & 96 & 0.38 [0.28, 0.50] & 0.32 [0.27, 0.41] & 0.40 [0.28, 0.53] & 0.33 \\
\bottomrule
\end{tabular}

\end{table*}

\paragraph{Sensitivity.}
\Cref{tab:sens} splits the wait of \cref{tab:latency} by the edit a build follows and by its place in its session.
Every class of edit gains. On 32 cores, the ratio runs from \NUM{SensZoneS}{}$\times$ after an edit of one scene's
script, which static impact confines to that scene's window, to \NUM{SensDataS}{}$\times$ after an edit of a data file,
which more frames read. The two rare classes gain as well. The ratio is \NUM{SensStyleS}{}$\times$ after an edit of the
style and \NUM{SensNoneS}{}$\times$ for a build of an unchanged version, whose film both systems return whole. A
session's first build has an empty cache, and on it the agent waits \NUM{SensFirstS}{}$\times$ as long, through the
verdict first alone. From a session's sixth build on, the agent waits \NUM{SensLateS}{}$\times$ as long. On builds
that deliver a film, the ratio holds across film lengths. By thirds of length on 32 cores, it is \NUM{SensFilmShortS}{}
in the shortest third, \NUM{SensFilmMidS}{} in the middle third and \NUM{SensFilmLongS}{} in the longest.
For builds that fail a check, the gain grows with the length, from \NUM{SensNoFilmShortS}{}$\times$ in the shortest
third to \NUM{SensNoFilmLongS}{}$\times$ in the longest. Over these thirds, \rebuild's checks take
\NUM{SensCheckShortBase}{} to \NUM{SensCheckLongBase}{}\,s, while the checks under \sys, which reuse their results,
take \NUM{SensCheckLongOurs}{}\,s in the longest third.

\begin{table}[t]
\centering
\caption{The wait under \sys relative to \rebuild by the edit a build follows and by its place in its session
(ratios of total waits).}
\label{tab:sens}
\footnotesize
\begin{tabular}{lrcc}
\toprule
Builds & $n$ & 32 cores & 8 cores \\
\midrule
First of a session (cache empty) & 70 & 0.56 & 0.60 \\
\midrule
\multicolumn{4}{l}{\emph{After an edit of}} \\
\quad one scene's script & 47 & 0.62 & 0.64 \\
\quad narration lines & 171 & 0.66 & 0.66 \\
\quad files added or removed & 277 & 0.62 & 0.66 \\
\quad a shared script & 74 & 0.66 & 0.72 \\
\quad a data file & 52 & 0.68 & 0.76 \\
\quad the style & 4 & 0.55 & 0.61 \\
\quad nothing (built again) & 7 & 0.29 & 0.19 \\
\midrule
\multicolumn{4}{l}{\emph{By place in the session}} \\
\quad second & 70 & 0.55 & 0.66 \\
\quad third to fifth & 204 & 0.67 & 0.72 \\
\quad sixth and later & 358 & 0.64 & 0.65 \\
\midrule
Every build & 702 & 0.64 & 0.67 \\
\bottomrule
\end{tabular}

\end{table}

\paragraph{What drives the gain, within runs.}
In the run of the last block of \cref{tab:ablation}, a session's change in wait follows the frames it saves. The
slope is \NUM{LawFactSlope}{} (CI \NUM{LawFactSlopeCI}{}) from \rebuild to reuse alone and \NUM{LawFactSchedSlope}{}
(CI \NUM{LawFactSchedSlopeCI}{}) from the verdict first alone to \sys. With the verdict first off or on, this is the
slope of \cref{fig:law}. Per stage (\cref{tab:stages}), reuse alone cuts both the checks, which reuse their samples, and
the capture. Reuse alone cuts the capture to \NUM{StageCaptureReuse}{}$\times$ \rebuild's time while rendering
\NUM{AblReuseFrames}{}\% fewer frames. The verdict first alone cuts the checks to \NUM{StageChecksSched}{}$\times$.
\sys cuts the checks to \NUM{StageChecksOurs}{}$\times$ and the capture to \NUM{StageCaptureOurs}{}$\times$.
\begin{table}[h]
\centering\footnotesize\setlength{\tabcolsep}{4pt}
\caption{The wait, the frames rendered and the time per stage in the run of the last block of \cref{tab:ablation}:
\rebuild's amounts per build, then each arm relative to them. A stage's time sums the wall-clock time of its operations,
which can overlap. The keys' time is given as a share of \rebuild's capture time.}
\label{tab:stages}
\resizebox{\columnwidth}{!}{\begin{tabular}{lrrrrrr}
\toprule
 &  &  & \multicolumn{3}{c}{\rebuild with one mechanism} &  \\
\cmidrule(lr){4-6}
Stage & \rebuild{} (per build) & \sequential & reuse & verdict first & fast checks & \sys \\
\midrule
Wait & 15.4\,s & 1.51 & 0.94 & 0.69 & 0.82 & 0.64 \\
Frames & 1{,}399 & 1.00 & 0.66 & 1.00 & 1.00 & 0.65 \\
\midrule
Checks & 10.8\,s & 0.78 & 0.89 & 0.59 & 0.38 & 0.26 \\
Keys (of capture) & -- & 0 & 0.05 & 0 & 0 & 0.06 \\
Capture & 7.1\,s & 1.25 & 0.87 & 1.07 & 1.06 & 0.91 \\
\bottomrule
\end{tabular}
}
\end{table}
\paragraph{One mechanism removed.}
Every mechanism carries part of the gain. On 32 cores, over \NUM{AblTopics}{} topics, the agent waits
\NUM{AblNoVerdict}{}$\times$ as long under \sys without the verdict first as under \sys (CI \NUM{AblNoVerdictCI}{}).
Without the fast checks, the agent waits \NUM{AblNoFast}{}$\times$ as long (CI \NUM{AblNoFastCI}{}). Without reuse, it
waits \NUM{AblNoReuse}{}$\times$ as long (CI \NUM{AblNoReuseCI}{}), with \NUM{AblNoReuseCpu}{}$\times$ the CPU. This
last arm is the scheduling part alone, and under it the agent waits \NUM{AblNoReuseBase}{}$\times$ as long as under
\rebuild (CI \NUM{AblNoReuseBaseCI}{}).

\par\noindent\textbf{Across sessions.}
The agent waits less in sessions that render fewer frames again (\cref{fig:law}a). The least-squares line is
$\text{wait}=\NUM{LawFitA}{}+\NUM{LawFitB}{}\cdot\text{frames}$, with both quantities relative to \rebuild (slope 95\%
CI \NUM{LawFitBCI}{}). The ablation separates what reuse and scheduling each contribute.
Opus~5.5's sessions on the agents' nine topics ($\dagger$), whose edits keep the most frames, gain most. The gain is
there from a session's first build, whose frame store is empty, on 32 and on 8 cores (\cref{tab:sens}).

\paragraph{CPU allocation.}
\begin{figure}[t]
\centering
\begin{tikzpicture}
\begin{axis}[
  width=\columnwidth, height=3.2cm, xmode=log, log basis x=2, xtick={4,8,16,32}, xmin=3.4, xmax=37.6,
  table/col sep=comma, xticklabels from table={figures/data/scaling.csv}{label},
  ymin=0, xlabel={cores (median render workers)}, ylabel={wait per build (s)},
  xlabel style={font=\sffamily\small, color=cInk}, ylabel style={font=\sffamily\small, color=cInk},
  ticklabel style={font=\sffamily\small, color=cInk}, axis line style={color=cGray}, tick style={color=cGray},
  axis x line*=bottom, axis y line*=left,
  legend style={at={(0.98,0.98)}, anchor=north east, draw=none, font=\sffamily\small}, legend cell align=left,
  error bars/error bar style={line width=0.5pt}, error bars/error mark options={rotate=90, mark size=1.4pt, line width=0.5pt},
]
\addplot[color=cGray, thick, mark=*, mark size=1.5pt, mark options={draw=white, line width=0.3pt}, error bars/.cd, y dir=both, y explicit]
  table[x expr=0.94*\thisrow{cores}, y=base, y error plus expr=\thisrow{base_hi}-\thisrow{base},
        y error minus expr=\thisrow{base}-\thisrow{base_lo}, col sep=comma] {figures/data/scaling.csv};
\addplot[color=cA, thick, mark=*, mark size=1.5pt, mark options={draw=white, line width=0.3pt}, error bars/.cd, y dir=both, y explicit]
  table[x expr=1.06*\thisrow{cores}, y=ours, y error plus expr=\thisrow{ours_hi}-\thisrow{ours},
        y error minus expr=\thisrow{ours}-\thisrow{ours_lo}, col sep=comma] {figures/data/scaling.csv};
\legend{\rebuild, \sys}
\end{axis}
\end{tikzpicture}%
\vspace{-2mm}
\caption{Mean wait per build against the session's cores, \rebuild and \sys; bars are 95\% intervals over topics.}
\label{fig:scaling}
\end{figure}
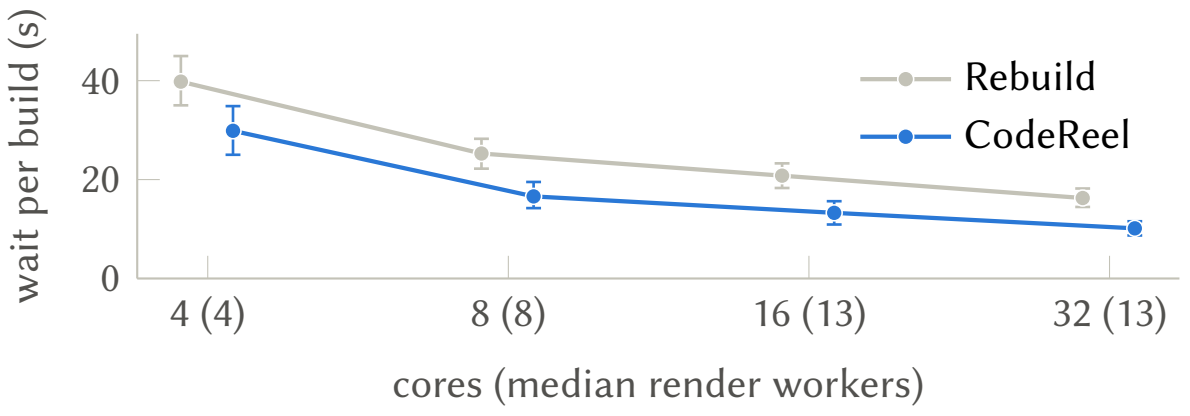
With 4 to 32 cores on \NUM{ScaleTopics}{} topics (\cref{fig:scaling}), the render's workers follow the cores until
the rule that sizes them by the film's length caps them, at 13 workers on these films. The cores beyond serve the
checks and the encoders. The wait under both systems still falls from 16 to 32 cores. The advantage of \sys grows with
the cores. The agent waits \NUM{ScanW}{}$\times$ as long under \sys as under \rebuild on 4 cores,
\NUM{ScanM}{}$\times$ on 8, \NUM{ScanMSixteen}{}$\times$ on 16 and \NUM{ScanS}{}$\times$ on 32, as the verdict first
spends on the checks the CPU that reuse frees. On the \NUM{TopicsM}{} timed topics of \cref{tab:results}, the agent
waits \NUM{MSixteenWait}{}$\times$ as long under \sys as under \rebuild on 16 cores (CI \NUM{MSixteenWaitCI}{}).

\par\noindent\textbf{Think time and work ahead.}
Work ahead in the agent's think time shortens the wait further. Replayed with the recorded think time, the
agent waits \NUM{ThinkWaitL}{}$\times$ as long under \sys as under \rebuild on 8 cores (CI
\NUM{ThinkWaitLCI}{}) and \NUM{ThinkWaitLS}{}$\times$ on 32 (CI \NUM{ThinkWaitLSCI}{}). On 8 cores, this wait is
\NUM{ThinkWaitPre}{}$\times$ the wait with work ahead off (CI \NUM{ThinkWaitPreCI}{}). With work ahead, builds render
\NUM{ThinkFramesCut}{}\% fewer frames, and the wait for films falls by \NUM{ThinkFilmCut}{}\%.

\paragraph{Starting the render early.}
We measure the early render on 32 cores, over \NUM{SSevTopics}{} topics, in a four-arm run with every film gated
(\NUM{SSevGate}{}). In this run, the rule starts the render early on \NUM{EarlyShareS}{}\% of the builds, and the
render keeps \NUM{IdleKeptS}{}$\times$ its speed alongside the checks. When the render never starts early, the agent
waits \NUM{SSevNever}{}$\times$ as long as under \sys (CI \NUM{SSevNeverCI}{}), and \NUM{SSevNeverFilm}{}$\times$ as
long on builds that deliver a film (CI \NUM{SSevNeverFilmCI}{}). \sys speeds up its key evaluation in two ways, by
drawing on contexts reset in place and by evaluating keys in the capture workers instead of in a pass before the
capture. Without these two speed-ups, the agent waits \NUM{SSevVFour}{}$\times$ as long as under the full \sys on these
topics (CI \NUM{SSevVFourCI}{}), and \NUM{SSevVFourBase}{}$\times$ as long as under \rebuild (CI
\NUM{SSevVFourBaseCI}{}).

\paragraph{Co-located agents.}
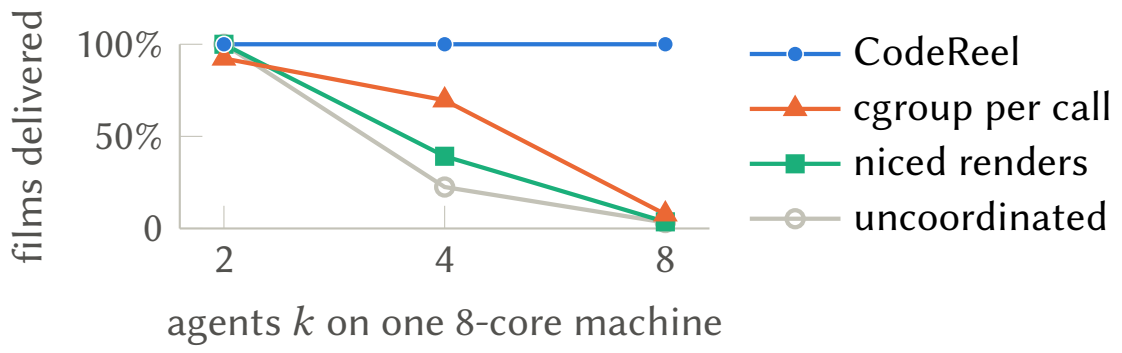
\begin{figure}[t]
\centering
\begin{tikzpicture}
\begin{axis}[
  width=0.60\columnwidth, height=2.8cm, xmode=log, log basis x=2, xtick={2,4,8}, xticklabels={2,4,8}, ymin=0, ymax=1,
  ytick={0,0.5,1}, yticklabels={0,50\%,100\%}, xlabel={agents $k$ on one 8-core machine}, ylabel={films delivered},
  xlabel style={font=\sffamily\small, color=cInk}, ylabel style={font=\sffamily\small, color=cInk},
  ticklabel style={font=\sffamily\small, color=cInk}, axis line style={color=cGray}, tick style={color=cGray},
  axis x line*=bottom, axis y line*=left, clip=false, table/col sep=comma, table/x=k,
  legend style={at={(1.04,0.5)}, anchor=west, draw=none, font=\sffamily\small, legend columns=1, row sep=-0.5pt},
  legend cell align=left, reverse legend,
]
\addplot[color=cGray, thick, mark=o, mark size=1.6pt, mark options={fill=white}] table[y=none] {figures/data/colocate.csv};
\addplot[color=cC, thick, mark=square*, mark size=1.4pt] table[y=nice] {figures/data/colocate.csv};
\addplot[color=cB, thick, mark=triangle*, mark size=1.9pt] table[y=cgroup] {figures/data/colocate.csv};
\addplot[color=cA, thick, mark=*, mark size=1.5pt, mark options={draw=white, line width=0.3pt}] table[y=ours] {figures/data/colocate.csv};
\legend{uncoordinated, niced renders, cgroup per call, \sys}
\end{axis}
\end{tikzpicture}%
\vspace{-2mm}
\caption{Films delivered with $k$ agents sharing one 8-core machine.}
\label{fig:colocate}
\end{figure}
We run $k$ agents on one machine of 8 cores with 16\,GB, in \NUM{ColoRunsPerK}{} runs per $k$, each of which draws
the $k$ sessions and their order at random (\cref{fig:colocate}). Whether an agent gets its films depends on memory.
Under uncoordinated execution, the agents' renders together outgrow the machine. At $k=4$ and $8$, the kernel's OOM
killer ended \NUM{ColoKilledNone}{} uncoordinated runs and \NUM{ColoKilledNice}{} runs with niced renders. One control
group per call~\cite{agentcgroup} keeps the runs alive by killing calls instead, with \NUM{ColoCgroupKills}{} kills at
$k=4$ and $8$, and \NUM{ColoCgroupLost}{} films are lost. Under \sys, with one ledger of the machine's cores and memory
and one shared executor (\cref{sec:colocate}), every run delivers every film that the recording delivered. These
\NUM{ColoOursFilms}{} films each pass the gate, and no call is killed. From $k=4$, only \sys delivers every film
(\cref{fig:colocate}).

\subsection{Exactness and tightness of reuse}

Over the \NUM{PairFrames}{} frames of the reuse study's \NUM{Pairs}{} pairs, \sys reuses no frame whose pixels
changed and renders close to the frames whose pixels are new. Every other scheme renders more or reuses stale frames
(\cref{tab:exact}). \sys renders \NUM{RenderedPct}{}\% of the frames, which is \NUM{Tightness}{}$\times$ the frames
whose pixels are new. Complete keys on whole scenes and on narration lines are exact as well. Complete keys on whole
scenes, which bound what per-call and per-scene caches~\cite{manim,fourier} can reuse, render \NUM{IdealScenePct}{}\%
of the frames, and complete keys on narration lines render \NUM{IdealLinePct}{}\%. Of the frames an edit changes,
\NUM{ChangedNearPct}{}\% keep an SSIM above 0.99 to the frame they replace, and \NUM{ChangedNearTightPct}{}\% keep an
SSIM above 0.999. The similarity rule of \cref{tab:exact} therefore leaves \NUM{SimStrictStalePct}{}\% of all frames
stale. These frames miss the agent's edit (\cref{fig:gallery}, bottom) at a median SSIM of \NUM{StaleSsim}{} to the
correct frame (see the similarity rule, below). No frame \sys reuses is stale, here or in \NUM{RandEdits}{} random
edits. A key audit checks every key against a lossless capture of its frame, over \NUM{AuditFrames}{} frames, of which
\NUM{AuditKeyedPct}{}\% are keyed. No key stands for two pictures, and every kept key audited equals its fresh
evaluation.

\paragraph{Stale frames per scheme.}
\Cref{tab:exact-full} extends \cref{tab:exact} with the frames that each scheme reuses although their pixels changed,
and with the input hash scheme.

\begin{table}[t]
\centering\footnotesize\setlength{\tabcolsep}{3pt}
\caption{Frames rendered and stale frames per reuse scheme, on the test split's adjacent versions and on time-shifting
narration edits (cv.js).}
\label{tab:exact-full}
\begin{tabular}{lrrrr}
\toprule
 & \multicolumn{2}{c}{Adjacent versions} & \multicolumn{2}{c}{Time-shifting edits} \\
\cmidrule(lr){2-3}\cmidrule(lr){4-5}
Method & Rendered & Stale & Rendered & Stale \\
\midrule
Re-render all, input hash & 100.0\% & 0 & 100.0\% & 0 \\
Scene windows & 90.3\% & 22{,}563 & 100.0\% & 0 \\
Read sets & 83.8\% & 9{,}704 & 53.2\% & 7{,}228 \\
Kamishibai-style, typical~\cite{kamishibai} & 63.5\% & 131{,}408 & 54.0\% & 6{,}121 \\
Kamishibai-style, careful & 89.0\% & 15{,}305 & 99.8\% & 0 \\
Manim-style~\cite{manim} & 88.0\% & 3{,}642 & 99.8\% & 0 \\
Similarity, threshold 0.1~\cite{teacache} & 59.7\% & 71{,}329 & -- & -- \\
Per scene, complete keys & 80.2\% & 0 & 81.1\% & 0 \\
Per line, complete keys & 71.1\% & 0 & 64.5\% & 0 \\
\sys & 63.2\% & 0 & 43.9\% & 0 \\
\midrule
Pixel truth & 60.2\% & 0 & 43.0\% & 0 \\
\bottomrule
\end{tabular}

\end{table}
\paragraph{Evidence.}
We compare incremental renders with renders from scratch, frame by frame, on \NUM{RandEdits}{} random edits of seven
kinds on four films (\NUM{RandKinds}{}). Every edit rendered, and every frame of every edit equals the render from
scratch. The same holds over the adjacent versions of the test split (\cref{tab:exact}).

\paragraph{Exports.}
Every one of the \NUM{Pairs}{} pairs and every random edit was exported and compared frame by frame. All
\NUM{PairFrames}{} frames of the pairs, reused or rendered, equal a render from scratch.

\paragraph{The similarity rule.}
\newcommand\stalessim{0.993}\newcommand\stalepsnr{34.2}\newcommand\stalechanged{0.37}
At the threshold of \cref{tab:exact}, the similarity rule leaves \NUM{SimStrictStalePct}{}\% of all frames stale. Its
stale frames look right to image metrics, at a median PSNR of \NUM{StalePsnr}{}\,dB and a median SSIM of
\NUM{StaleSsim}{}, because an agent often changes a few words. In one example, a caption was rewritten from ``the weak
machine is about 4$\times$ slower'' to ``the weak machine is about 4$\times$ slower than the strong machine''
(translated from Chinese), and the frame the rule keeps has an SSIM of \stalessim{} to the correct frame.
Whatever the rule, many of the frames an edit changes stay near the frames they replace. Of the frames that differ
from the previous version's frame at the same time, \NUM{ChangedNearPct}{}\% have an SSIM above 0.99 to that frame and
\NUM{ChangedNearTightPct}{}\% have one above 0.999, and the median SSIM is \NUM{ChangedNearMedian}{}. A rule that
reused every frame above 0.99 would leave \NUM{ChangedNearAllPct}{}\% of all frames stale. The shares are measured on
thirty changed frames drawn from each pair of \cref{tab:exact}, rendered in both versions and weighted by the pair's
changed frames.
\Cref{tab:eps} gives the similarity rule at thresholds below that of \cref{tab:exact}, on \NUM{EpsPairs}{} pairs of
the test split drawn at random. The smallest mean difference that two different $32\times18$ thumbnails can have is
$1/576$. At \NUM{EpsMin}{}, below that value, the rule reuses a frame only when its thumbnail is identical to the
previous version's. A stricter threshold makes the rule render more frames and reuse fewer stale ones. At the strictest
threshold, the rule renders \NUM{EpsMinPct}{}\% of the frames, more than the \NUM{EpsTruthPct}{}\% whose pixels
changed, and still reuses \NUM{EpsMinStale}{} stale frames in \NUM{EpsMinStalePairs}{} pairs. Matching each frame to
the nearest thumbnail of the previous version lets the rule reuse a frame moved in time. At the strictest threshold,
this matching reuses \NUM{EpsMinStaleNn}{} stale frames in \NUM{EpsMinStaleNnPairs}{} pairs. Two frames with the same
thumbnail can differ, so no threshold on the thumbnail makes the rule exact.

\begin{table}[t]
\centering
\caption{The similarity rule at stricter thresholds, on \NUM{EpsPairs}{} random pairs: frames rendered and stale
frames (pairs with one), reusing the previous version's frame at the same time or the one with the nearest thumbnail.}
\label{tab:eps}
\footnotesize
\setlength\tabcolsep{4pt}
\begin{tabular}{lrrrr}
\toprule
 & \multicolumn{2}{c}{Same time} & \multicolumn{2}{c}{Nearest} \\
\cmidrule(lr){2-3}\cmidrule(lr){4-5}
Threshold & Rendered & Stale & Rendered & Stale \\
\midrule
Pixels changed & 57.2\% & 0 & 47.7\% & 0 \\
\midrule
0.1 & 51.1\% & 4{,}146 (14) & 39.4\% & 15{,}216 (18) \\
0.05 & 52.6\% & 3{,}441 (12) & 41.2\% & 14{,}175 (18) \\
0.01 & 62.9\% & 108 (7) & 57.4\% & 6{,}648 (18) \\
0.001 & 66.7\% & 14 (4) & 69.6\% & 1{,}804 (17) \\
\bottomrule
\end{tabular}

\end{table}

\par\noindent\textbf{Time-shifting edits.}
\sys reuses the frames that a time-shifting edit moves. Such a move defeats reuse by time and by scene window. On
\NUM{ShiftEdits}{} narration edits to four cv.js films (\cref{tab:exact}, right), \sys renders
\NUM{ShiftTabOursPct}{}\% of their frames, against \NUM{ShiftTabTruthPct}{}\% whose pixels are new, and reuses no stale
frame. Read sets render more and reuse \NUM{ShiftTabReadStale}{} stale frames. On two Remotion films with eight such
edits each, \NUM{ShiftRxTruth}{} distinct pictures are new. \sys renders \NUM{ShiftRxOurs}{} frames there, and a cache
looked up by time renders \NUM{ShiftRxPos}{}.

\paragraph{Manim CE's own cache.}
On one scene of eight animation calls, Manim CE's cache re-renders every frame of each call an edit reaches. This is
the granularity of Manim-style per-call caching in \cref{tab:exact}. Manim CE re-renders from 16 of 106 frames after a
caption edit to every frame after a change to the text every call draws, and it shows no stale frame.
The stale frames of Manim-style per-call caching lie in scenes whose own files did not change but which call drawing
code from another scene's changed file. A per-call key over a call's own file, the shared scripts and its data cannot
see that code, which immediate-mode programs run every frame. Manim instead hashes the objects themselves.

\par\noindent\textbf{The cost of each edit.}
On \NUM{EditWithinFivePct}{}\% of the pairs, \sys renders within 5 percentage points of the frames whose pixels are
new, with a median excess of \NUM{EditExcessMed}{} points (\cref{fig:law}b). Edit-time
methods~\cite{lazydiffusion,sige} aim for such a cost, close to what the edit changes. Manim-style per-call caching
renders every frame of each call an edit reaches.

\subsection{Real agents in the loop}
\paragraph{Real-agent runs.}
Every session ran on 8 cores with cv.js programs and the scripted user of the registered protocol. The scripted user
asks for each round's changes and accepts the film. Claude Code ran Claude Opus~5.5, and Codex CLI ran GPT-6.1. The
Wait and Film ratios of \cref{tab:agents} are geometric means over pairs of per-session medians. The table's Total column gives each pair's ratio of total waits, the measure of the replays. Replayed under the
protocol of \cref{sec:experiments}, the sessions that Claude Code recorded live on topics never used in development
show the same gain. The agent waits \NUM{UntouchedWaitS}{}$\times$ as long as under \rebuild on 32 cores and
\NUM{UntouchedWaitM}{}$\times$ on 8.

\paragraph{Film quality of the real-agent pairs.}
Under the registered scoring protocol, Claude Opus judges the delivered films and lists the faults of each session's
final film. We report the paired differences, \sys\ minus \rebuild. On the \NUM{RPairs}{} pairs whose topics
development never touched, the final films' faults change by \NUM{RQualPrimaryFaults}{} (95\% CI
\NUM{RQualPrimaryFaultsCI}{}). For the \NUM{RQualCodexPairs}{} pairs through Codex CLI, the faults change by
\NUM{RQualCodexFaults}{}. The judge sees each film's contact sheets, last frame, narration text and audio levels under a
name drawn from a content hash, so the judge does not know which system built the film. Claude Code's 26 pairs are ten
topics run twice and six topics of Code2Video's benchmark~\cite{code2video} run once. Codex CLI's pairs are those six
topics run twice. The two sessions of each pair ran back to back on the same cores, in an order that alternates between
repetitions. Every interval resamples topics with all their pairs. Each final film that a build under \sys delivered
was rendered again from scratch with the session's own code, and the two films decode to the same frames.

\subsection{Ablation study}

The two parts shorten the wait in different ways. For the last block of \cref{tab:results}, we replay \rebuild with
one mechanism at a time on 32 and 8 cores over \NUM{AblTopics}{} topics, without think time. Reuse removes work. Added
alone, reuse renders \NUM{AblReuseFrames}{}\% fewer frames with \NUM{AblReuseCpu}{}$\times$ the CPU. The agent then
waits \NUM{AblReuse}{}$\times$ as long on 32 cores and \NUM{AblReuseM}{}$\times$ on 8, where rendering fills the
machine. The verdict first, which runs the checks in parallel and the render alongside them, spends CPU to shorten the wait. With
the verdict first alone, the agent waits \NUM{AblSched}{}$\times$ as long on 32 cores and \NUM{AblSchedM}{}$\times$ on
8. On failing builds, it waits \NUM{AblSchedFail}{}$\times$ as long on 32 cores and \NUM{AblSchedFailM}{}$\times$ on
8. The verdict first alone uses \NUM{AblSchedCpu}{}$\times$ the CPU on 32 cores and \NUM{AblSchedCpuM}{}$\times$ on 8.
With every mechanism, the agent waits \NUM{AblOurs}{}$\times$ as long under \sys on 32 cores and
\NUM{AblOursM}{}$\times$ on 8, and \sys uses \NUM{AblOursCpu}{}$\times$ the CPU on 32 cores and
\NUM{AblOursCpuM}{}$\times$ on 8. Reuse thus returns the CPU that the verdict first spends. With the fast checks alone,
the agent waits \NUM{AblFast}{}$\times$ as long on 32 cores and \NUM{AblFastM}{}$\times$ on 8. Starting the render
early adds to the gain of the verdict first. When the render never starts early, the agent waits
\NUM{SSevNever}{}$\times$ as long as under \sys on 32 cores (CI \NUM{SSevNeverCI}{}). Without static impact, \sys
evaluates \NUM{PartKeysImpact}{}$\times$ as many keys, and without values read, it evaluates
\NUM{PartKeysValue}{}$\times$ as many (see the parts of reuse, below).

\paragraph{The parts of reuse.}
We turn one part of reuse off at a time on \NUM{AblPairs}{} pairs of \cref{tab:exact}. With static impact or values
read off, the same frames are rendered. The keys evaluated rise to \NUM{PartKeysImpact}{}$\times$ with static impact
off and to \NUM{PartKeysValue}{}$\times$ with values read off. Text-baseline snapping saves \NUM{ExtraNormtext}{}
unchanged frames. Keys that hold a frame's position in time cost \NUM{PositionalLines}{}$\times$ the frames on
narration edits. We also time \sys with one part off at a time on four topics, with every film gated
(\NUM{SwGates}{}). Relative to the full \sys, the agent waits \NUM{SwImpact}{}$\times$ as long with static impact off,
\NUM{SwValue}{}$\times$ with values read off and \NUM{SwNormtext}{}$\times$ with text-baseline snapping off.

\section{Related work in full}
\label{sec:related-full}
\paragraph{Code-to-video agents and benchmarks.}
Agents make explanatory videos by writing Manim, web or slide programs and refining them with a renderer in the
loop~\cite{code2video,theoremexplain,omnimanim,reelsrsi,paper2video}. Agents also drive physics simulators and
renderers through code~\cite{moregen}, mix programs with generative models~\cite{preacher}, and program slides and 3D
scenes in the same loop~\cite{autopresent,blenderalchemy,blendergym}. VISTA~\cite{vista} refines generated videos,
LottieGPT generates vector animations as tokens~\cite{lottiegpt}, and the Scene Language represents scenes as
programs~\cite{scenelanguage}.
These works improve what the agent writes and how its output is judged, and some measure tokens and
time~\cite{code2video,omnimanim}. None goes beyond its framework's per-call cache or targets the latency of a verdict.
Benchmarks grade the result, and some also grade cost or trajectories~\cite{blendergym,compass,timelinebench}.
Code2Video-Compass~\cite{compass}, on the same web and Remotion stack, generates a video in one call. Agent traces
record tool latencies~\cite{tracelab} and are replayed to benchmark serving~\cite{agentreplay,agentperfbench}.
\bench's replay instead executes the calls.

\paragraph{Reuse.}
Training-free caches reuse or forecast features between denoising
steps~\cite{deepcache,blockcache,pab,fastercache,teacache,adacache,taylorseer}. Edit-time methods compute on the edited
region~\cite{sige,lazydiffusion,fisedit,hetcache} or reuse unchanged regions across editing turns~\cite{regioncache}.
Serving, graphics and video models skip or reproject work for similar inputs, frames or
pixels~\cite{streamdiffusion,nirvana,rendercache,reprojection,skipconv,deltacnn,eventful}. All of these methods let
their output differ from the full computation's.
Production tools and UI toolkits reuse operator outputs, frames and display lists~\cite{nuke,aegpc,blinkpaint,flutter}.
Closest to us, Manim keys each animation call by its inputs~\cite{manim}. Kamishibai fingerprints each frame's page and
reuses the previous run's frame at the same index~\cite{kamishibai}. HyperFrames repeats the frames that its timeline
predicts to be static within one render~\cite{hyperframes}. Fourier, a video framework built for agents, fingerprints
sources, assets and timing to reuse unchanged scenes, layers and speech, also after a layer moves in
time~\cite{fourier}.
\sys keys a frame by its evaluated drawing commands, normalized to the rasterizer, so frames of imperative code with
one global clock and no call or layer boundaries are found again in any version, at any time. Like build
systems~\cite{buildsalacarte,riker} and incremental computation~\cite{hammer2014adapton,hammer2015adapton}, \sys keeps
its results consistent with a build from scratch. Its keys, however, come from what a program draws and not from the
files the program reads.

\paragraph{Scheduling around agents.}
Test prioritization runs likely failures first~\cite{rothermel2001apfd,prado2020ntr}. Agent controllers weigh cheap
diagnostics against costly verification~\cite{bayesctl}. Agent serving places KV caches around tool
calls~\cite{mori,askthetool}, shares GPU servers among agents~\cite{justitia}, prewarms sandboxes~\cite{specbox} and
controls CPU and memory per tool call by declared intent~\cite{agentcgroup}. Agent systems also run independent tool
calls in parallel~\cite{llmcompiler} and start a tool while the model is still decoding~\cite{conveyor}. Speculative
agent systems guess the agent's next action and keep a guess only when the guess matches the action
taken~\cite{specactions}. Build services schedule by a priority that users set~\cite{buildbuddy}.
\sys schedules the operations of a video build by what the agent is blocked on, within the build and across the agent's
think time. It admits the work of agents sharing a machine against one ledger of the machine's cores and memory. The
work ahead of \sys needs no rollback, since a frame stored under its key is exact for any later version that draws it.

\end{document}